\documentclass[amsmath,amssymb,twocolumn,notitlepage,nofootinbib,superscriptaddress,floatfix,eprint]{revtex4-2}
\usepackage{lineno}
\usepackage[utf8]{inputenc}
\usepackage{graphicx}
\usepackage{braket}
\usepackage{multirow}
\usepackage{booktabs}
\usepackage{hyperref}
\usepackage{xcolor}
\usepackage{float}
\usepackage{bm}
\usepackage{soul}
\usepackage{mathtools}
\usepackage{makecell}

\newcommand{\changes}{\textcolor{blue}}

\begin{document}

\title{Experimental neuromorphic computing based on quantum memristor}

\author{Mirela Selimovi\'{c}}
\email{mirela.selmovic@univie.ac.at}
\affiliation{University of Vienna, Faculty of Physics, Vienna Center for Quantum
Science and Technology (VCQ), Boltzmanngasse 5, Vienna 1090, Austria}
\affiliation{University of Vienna, Faculty of Physics, Vienna Doctoral School in Physics (VDSP), Boltzmanngasse 5, Vienna 1090, Austria}

\author{Iris Agresti}
\email{iris.agresti@univie.ac.at}
\affiliation{University of Vienna, Faculty of Physics, Vienna Center for Quantum
Science and Technology (VCQ), Boltzmanngasse 5, Vienna 1090, Austria}

\author{Michał Siemaszko}
\affiliation{Faculty of Mathematics, Informatics, and Mechanics, University of Warsaw, Stefana Banacha 2, 02-097 Warsaw, Poland}

\author{Joshua Morris}
\affiliation{University of Vienna, Faculty of Physics, Vienna Center for Quantum
Science and Technology (VCQ), Boltzmanngasse 5, Vienna 1090, Austria}
\affiliation{University of Vienna, Faculty of Physics, Vienna Doctoral School in Physics (VDSP), Boltzmanngasse 5, Vienna 1090, Austria}

\author{Borivoje Daki\'{c}}
\affiliation{University of Vienna, Faculty of Physics, Vienna Center for Quantum
Science and Technology (VCQ), Boltzmanngasse 5, Vienna 1090, Austria}
\affiliation{Institute for Quantum Optics and Quantum Information Sciences (IQOQI), Austrian Academy of Sciences, Boltzmanngasse 3, Vienna 1090, Austria}
\affiliation{QUBO Technology GmbH, 1090 Vienna, Austria}

\author{Riccardo Albiero}
\affiliation{Istituto di Fotonica e Nanotecnologie, Consiglio Nazionale delle Ricerche (IFN-CNR), piazza L. Da Vinci 32, 20133 Milano, Italy}

\author{Andrea Crespi}
\affiliation{Dipartimento di Fisica, Politecnico di Milano, piazza L. da Vinci 32, 20133 Milano, Italy}

\author{Francesco Ceccarelli}
\affiliation{Istituto di Fotonica e Nanotecnologie, Consiglio Nazionale delle Ricerche (IFN-CNR), piazza L. Da Vinci 32, 20133 Milano, Italy}

\author{Roberto Osellame}
\affiliation{Istituto di Fotonica e Nanotecnologie, Consiglio Nazionale delle Ricerche (IFN-CNR), piazza L. Da Vinci 32, 20133 Milano, Italy}

\author{Magdalena Stobińska}
\affiliation{Faculty of Mathematics, Informatics, and Mechanics, University of Warsaw, Stefana Banacha 2, 02-097 Warsaw, Poland}

\author{Philip Walther}
\email{philip.walther@univie.ac.at}
\affiliation{University of Vienna, Faculty of Physics, Vienna Center for Quantum
Science and Technology (VCQ), Boltzmanngasse 5, Vienna 1090, Austria}
\affiliation{Institute for Quantum Optics and Quantum Information Sciences (IQOQI), Austrian Academy of Sciences, Boltzmanngasse 3, Vienna 1090, Austria}
\affiliation{QUBO Technology GmbH, 1090 Vienna, Austria}

\begin{abstract}
Machine learning has recently developed novel approaches, mimicking the synapses of the human brain to achieve similarly efficient learning strategies. Such an approach retains the universality of standard methods, while attempting to circumvent their excessive requirements, which hinder their scalability. In this landscape, quantum (or quantum inspired) algorithms may bring enhancement.
However, high-performing neural networks invariably display nonlinear behaviours, which poses a challenge to quantum platforms, given the intrinsically linear evolution of closed systems.
We propose a strategy to enhance the nonlinearity achievable in this context, without resorting to entangling gates and report the first neuromorphic architecture based on a photonic quantum memristor.
In detail, we show how the memristive feedback loop enhances the nonlinearity and hence the performance of the tested algorithms. We benchmark our model on four tasks, a nonlinear function and three time series prediction. In these cases, we highlight the essential role of the quantum memristive element and demonstrate the possibility of using it as a building block in more sophisticated networks.

\end{abstract}

\maketitle

\section{Introduction}
In the last decades, machine learning has fundamentally changed the approach to solving a wide variety of problems, from everyday life to scientific scopes \cite{abdel-hamid_convolutional_2014, bakator_multimodal_2018, silver_mastering_2017, pierson_deep_2017, 755469, silver_mastering_2017, jumper_highly_2021, brown_language_2020}. 
In this context, artificial neural networks (ANN) have proven to be a pivotal tool, especially for tackling tasks where information needs to be extracted from high-dimensional data \cite{krizhevsky_imagenet_2017}. Nevertheless, a growing number of parameters must be trained as the tasks get more complex, with correspondingly larger training data, resulting in the so-called memory bottleneck \cite{mutlu_processing_2019}. This sparked the development of more dedicated hardware, i.e. neuromorphic architectures, which mimic the human brain \cite{ham_neuromorphic_2021}.
The main advantage of these models is their minimal learning requirements, which circumvents the aforementioned difficulty.
Acting as the poster-child of neuromorphic components is the memristor, a passive circuital component that was postulated in 1971 \cite{chua_memristor-missing_1971} and demonstrated in 2008 \cite{strukov_missing_2008}. Its main characteristic is that it retains memory of its past states, in the form of a hysteresis that is similar to neural synapses \cite{jo_nanoscale_2010, du_reservoir_2017, ham_neuromorphic_2021, xiao_review_2023, zhong_dynamic_2021}.

\begin{figure*}[ht]
    \centering
    \includegraphics[width=0.7\textwidth]{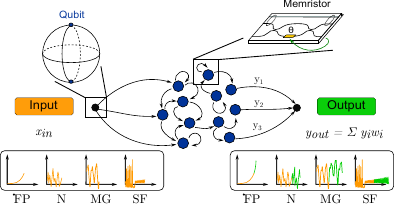}
    \caption{\textbf{Neuromorphic computing based on quantum memristor.} The basic principle of a quantum memristor is a tunable Mach-Zehnder interferometer, where its internal phase ($\theta$) is varied, according to the output registered at one of its output modes, through a feedback loop. The investigated hybrid quantum/classical models are implemented through three parts: input encoding (quantum), nonlinear reservoir (quantum) and linear regression, i.e. $y_{out}=\Sigma y_i w_i$ (classical. In our work, the encoding of classical variables in a one qubit state exploits the amplitude of the component parallel to $|0\rangle$, i.e. $x\rightarrow\sqrt{x}|0\rangle+\sqrt{1-x}|1\rangle$ and $x|0\rangle+\sqrt{1-x^2}|1\rangle$. The reservoir consists in a unitary operation, composed by the product of different rotations. These rotations allow us to achieve a nonlinear transformation of the input, due to the encoding that is nonlinear in the density matrix of the quantum state. The loop enhances the achievable nonlinearity, while implementing a short-term memory. The nonlinear reservoir is then followed by a linear regression model, which is the only part of the model that is trained, producing a weighted sum of the reservoir outputs. The four addressed tasks are the prediction of a smooth nonlinear function (FP) and three random time series: NARMA (N), Mackey-Glass (MG), and Santa Fe (SF).}
    \label{fig:conceptual-scheme}
\end{figure*}

\begin{figure*}[ht]
    \centering
    \begin{minipage}[c]{0.55\textwidth}
        \centering
        \includegraphics[width=\textwidth]{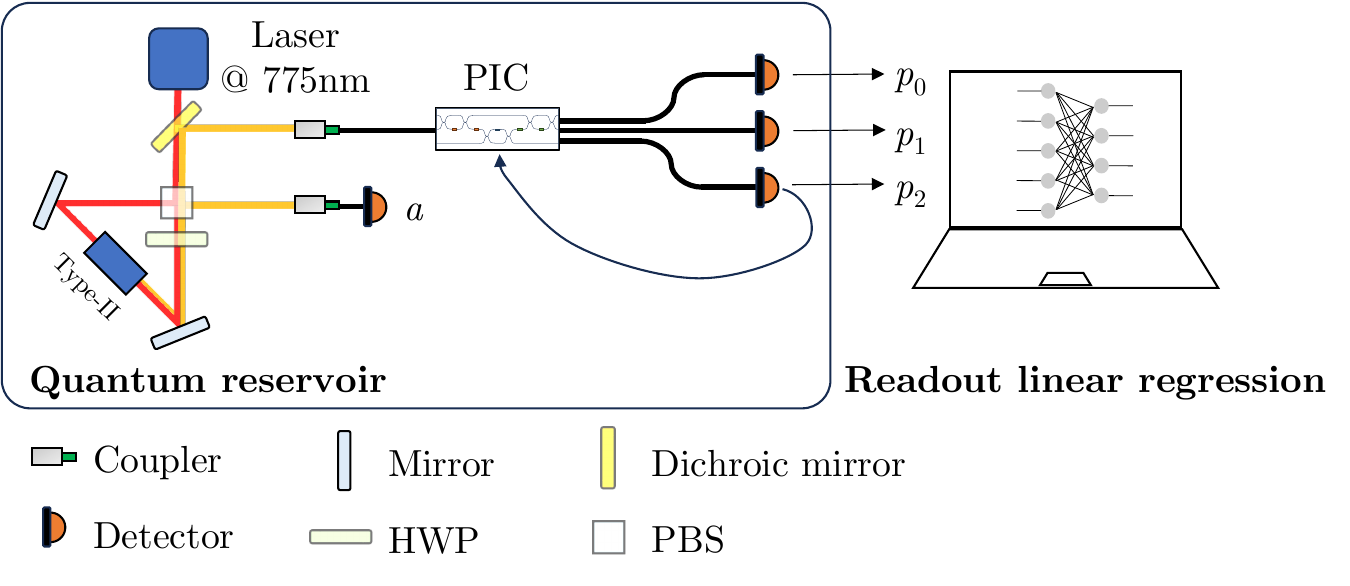}
        \vspace{1mm}
        
        \textbf{(a)}
    \end{minipage}
    \hfill
    \begin{minipage}[c]{0.4\textwidth}
        \centering
        \includegraphics[width=\textwidth]{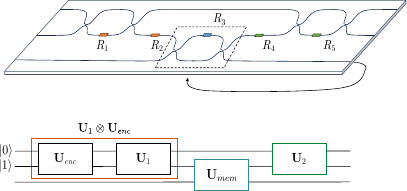}
        \vspace{1mm}
        
        \textbf{(b)}
    \end{minipage}

    \caption{\textbf{Experimental setup.} 
    \textbf{(a)} The experimental apparatus includes a source of single photon pairs pumped with a laser at 775~nm and generates degenerate photon pairs at 1550~nm. One photon is routed to the 3 input/output mode photonic integrated circuit (PIC) and the other photon is ancillary. Let us note that this circuit is implemented following the universal Reck architecture \cite{reck_experimental_1994}. At the outputs of the PIC, photon counts are registered by superconducting nanowire single photon detectors. After the full calibration of the chip, phases are set and modulated by applying a suitable voltage to the resistive elements, that act as tunable thermal phase shifters. The unitary operation is updated, conditioned on the statistics of previous runs. The output distribution is then fed into a linear regression model, which is the only part of the machine learning model that is trained. 
    \textbf{(b)} The optical circuit consists in 3 Mach-Zehnder interferometers and 5 tunable phases. The $R_i$ indicate the resistive elements, where $R_3$ is updated conditioned on the output of the third mode, implementing a memristive behaviour. Let us note that the state obtained tracing out the third output still preserves coherence (see Supplementary Material note I). The input is encoded in one qubit, through a unitary transformation $\mathbf{U}_{\text{enc}}$. The unitary transformations $\mathbf{U}_{1}$ and $\mathbf{U}_{2}$ are taken as hyperparameters of our model for the monomial functions. Note that we implement the product $\mathbf{U}_1 \cdot \mathbf{U}_{\text{enc}}$ with the first MZI. Then, $\mathbf{U}_{\text{mem}}$ amounts to the action of the memristor and is varied depending on previous outcomes. Also the coefficient of the linear feedback rule, according to which this happens, are hyperparameters, see Eq.~\eqref{eq:update_P}. Let us note that for the time series prediction tasks, $U_1 = U_3 = \mathbb{I}$, and the only hyperparameter is constituted by the memory decay $m_d$ (see Eq.~\eqref{eq:feedback_rule}).}
    \label{fig:experimental_setup}
\end{figure*}

Quantum computing introduced a related but somewhat orthogonal paradigm shift in computer science. The interest in this field stems, first, from the possibility of tackling non-classical processing tasks directly related to the investigation of quantum effects and, second, from the promise of outperforming standard algorithms for particular problems \cite{grover_fast_nodate, shor_polynomial-time_1997, montanaro_quantum_2016, harrow_quantum_2009}. However, for the latter, the only proven advantages were shown for tasks that are beyond the reach of even near-term state-of-the-art quantum computers or which have no known applications \cite{arute_quantum_2019}. 
Hence, a research line that has attracted a lot of interest arises from the combination of machine learning and quantum computation, that is, quantum machine learning (QML) \cite{dunjko_machine_2018, wittek2014quantum}. This is fueled by the hope that a computational advantage can be found on practically relevant tasks even when using relatively small-scale quantum information processing.
In this context, photonic apparatuses constitute a very promising platform, as they exhibit a vastly lower energy consumption compared to standard electrical ones \cite{mcmahon_physics_2023, hamerly2019large}, indicating the possibility of far more efficient and fast computational platforms.
However, when considering quantum processes, the natural way of implementing nonlinear evolutions (crucial for machine learning algorithms) consists in the (controlled) interaction among multiple systems, i.e. entangling gates. This kind of evolution is still governed by linear transformations, but it can be seen as nonlinear from a computational point of view. These can be deterministically achieved on several platforms, e.g. superconducting qubits or trapped ions, but the scalability is hindered by the rising of decoherence. In contrast, on photonic platforms, decoherence is less of an issue, but photonic nonlinear interactions are weak and hardly controllable. This implies, for example, that very high optical powers need to be used, which, besides the engineering challenges, can also negate the potential energy advantage coming from using optics instead of electronics. A possible way out is by resorting to optoelectronics \cite{hamerly2019large}, although it requires energy to convert light into electronic signals and back \cite{mcmahon_physics_2023}, and it slows down the whole process. This is the reason why, until now, this problem has been circumvented, either through a suitable encoding of the data, as in \cite{yin_experimental_2024, wanjura2024fully}, or by exploiting the nonlinearity coming from the measurement, as in \cite{suprano2024experimental, rausell2024programmable}. These approaches are all proven to be effective, but they present drawbacks. Indeed, the first cannot be used with quantum data and the second cannot implement any memory nor be used as a subroutine.

In this work, we propose a possible approach to solve this apparent conundrum and we implement the first instances of a neuromorphic computing protocol, i.e. reservoir computing (RC), using single photons, manipulated through an integrated circuit \cite{mujal_opportunities_2021}. The model (see Fig.~\ref{fig:conceptual-scheme}) may be seen as a building block for more complex networks and is composed of a nonlinear fixed reservoir (equipped with classical memory), followed by the readout unit, implementing a linear regression model, $y_{out} = \Sigma y_i w_i$, which is the only part that requires a training phase.
Our aim is to implement a quantum/classical hybrid algorithm, where the nonlinearity is generated by a quantum system, while the linear regression model is classical. To achieve this goal we leverage a particular choice for the input encoding and the exploitation of a novel device: the photonic quantum memristor \cite{spagnolo_experimental_2022}. This element has the two-fold effect of enhancing the nonlinearity of the model, while implementing a classical memory. This device is based on a Mach-Zehnder interferometer equipped with a feedback loop, and its behaviour is proven to be \textit{memristive}. Moreover, the impact of the feedback on the coherence of the output state was studied, both in a configuration featuring a single device and in a network of two independent quantum memristors, injected of correlated photon pairs \cite{ferrara_entanglement_2025}. However, its potential use for machine learning tasks was only theorized and its role was not deeply investigated, as the task proposed in the original manuscript, i.e. the recognition of MNIST hand-written digits \cite{li_deng_mnist_2012}, can be tackled effectively through linear regression and other simple estimators. Before scaling up to larger systems, we investigate the potentialities of one device, to understand what the minimal requirements are to tackle real-world machine learning tasks. To give a better estimation, we also carry out a comparison between our model and simple non-task-specific classical models, to quantify the resources they require to achieve a similar, or slightly better, performance.
Hence, here, we go beyond the previous study and apply this hardware to four machine learning tasks where nonlinearity is a crucial factor and cannot be omitted without compromising performance, highlighting the pivotal role of the quantum reservoir.
In all of the considered tasks, we show that the dynamics of the photonic quantum memristor enhances the performance, with respect to the case where no feedback loop is implemented.
Hence, the present work shows how a (photonic) quantum system can tackle tasks that were addressed with similar classical architectures \cite{du_reservoir_2017}, but with substantially less resources, paving the way to resource efficient machine learning models and to the possibility of exploring neuromorphic architectures applied to quantum tasks.

\section{Quantum memristor-based neuromorphic models}

The photonic quantum memristor can be modeled as a tunable Mach-Zehnder interferometer, as depicted in Fig.~\ref{fig:experimental_setup}. Then, its internal phase is updated according to a feedback rule that depends on the measurement outcome at one output mode, affecting the temporal dynamics of the device. One of its main features consists in the fact that, when tracing out the mode used for the feedback loop, the remaining reduced state may still be partially coherent relative to the joint state, as detailed in the Supplementary Material note I. Consequently, further quantum information processing is possible on the output.

We want to exploit this feedback mechanism to implement nonlinear operations on the input state and utilize a short-term memory, to achieve a proof-of-principle quantum reservoir computing (QRC).
In detail, we encode our classical inputs in quantum states that are then injected into a quantum reservoir. This usually consists of randomly connected nodes, whose internal time-based state make up its intrinsic memory and the historical nonlinearity that is necessary for the learning process (see the Supplementary Material note II for further details). In this framework, feedback-driven and (weak-) measurement-based architectures have been recently theoretically investigated \cite{kobayashi_feedback-driven_2024, mujal_time-series_2023, franceschetto_harnessing_2024}.
Our implementation features one physical node, realized through a Mach-Zehnder interferometer (MZI), which in concert with a feedback loop acts like a quantum memristor. Then, the randomness is given by unitary operations before and after the memristor itself. 
Hence, the quantum part of our model consists in the encoding of the classical data and reservoir. In particular, we exploit one-qubit states, encoded in the path degree of freedom of a single photon state, with a dual-rail scheme (see Fig. ~\ref{fig:experimental_setup}b).
The reservoir is then composed of a sequence of three unitaries over three spatial modes, i.e. $\textbf{U}_1$, $\textbf{U}_{mem}$ and $\textbf{U}_2$. 
At the end, to build a proper reservoir computing scheme, a readout unit, i.e. a classical linear regression model, follows.

\begin{figure}[t]
    \centering \includegraphics[width=\columnwidth]{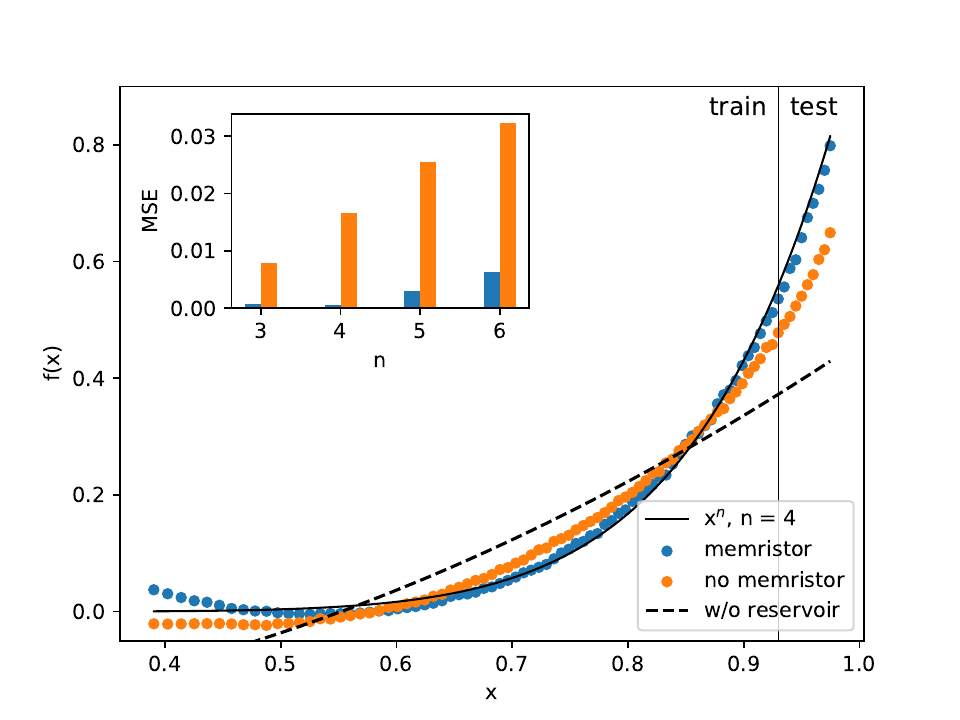}  \caption{\textbf{Nonlinear function prediction results.} The blue dots show the experimental results for the prediction of the power function $f_{t}(x)=x^n$ for $n=4$ (indicated by the solid black curve). The orange ones show the results for the same task, but without feedback loop. Inputs lower than 0.9 are used for training the model, while the other ones are used for test. The case with no feedback loop is on average 34\% less accurate than the one exploiting the memristor. Let us note that, to have a fair comparison, hyperparameter were optimized separately for the two cases. The dashed black line represents the case without the physical reservoir, i.e. where the inputs $x$ are directly injected to the readout linear regression. The loss function is defined as the mean squared error MSE between the target function $f_{t}(x)=y_{t}$ and the test data $y^k_{test}$, i.e. $y_{t}-y_{test}$. The inset shows the MSE for power function with $n=3,4,5,6$ in the QRC case (blue) and with no feedback loop case (orange).}
    \label{fig:final_fp}
\end{figure}

We test our model for four tasks, that rely on the two key features of our photonic platform, i.e. nonlinearity and short-term memory. These are a nonlinear function prediction/reproduction, where we exploit mostly the nonlinearity of the model, and three different time series predictions \textit{nonlinear Auto Regressive Moving Average, i.e. NARMA}, \cite{hochreiter_long_1997, atiya_new_2000} \textit{Mackey-Glass dynamic prediction} \cite{shahi2022prediction,fujii_harnessing_2017} and \textit{SantaFe laser intensity prediction} \cite{hubner1989dimensions, weigend1993results}, where short-term memory plays a pivotal role.

For the first task, the functions we aim at reproducing and then extrapolating are monomials:

\begin{equation}
    f(x) = x^n, \ n \in \mathbb{Z}
    \label{eq:f_2_predict}
\end{equation}

This choice is motivated by the interest in testing the maximal achievable nonlinear behaviour of our model and the role of the feedback loop towards its enhancement.
 
We encode the inputs as follows: $x \rightarrow \sqrt{x}|0\rangle + \sqrt{1-x}|1\rangle$. Note that this encoding is linear when performing a projective measurement in the eigenbasis of the Pauli operator $\sigma_Z$ \cite{govia_nonlinear_2022}, as $Tr(\sigma_Z \rho_x) = \sum_{i=0,1}(-1)^ip_i=2x-1$, where $p_i$ is the probability of getting the outcome $i$. In contrast, measurements in the $\sigma_X$ basis give an expectation value that is nonlinear in $x$, $Tr(\sigma_X \rho_x) = \sqrt{x}\sqrt{1-x}$. This nonlinearity can be enhanced by a proper choice of the rotations $\textbf{U}_1$ and $\textbf{U}_2$. Then, we pick a linear feedback rule which modifies $\textbf{U}_{mem}$. The feedback loop additionally enhances the nonlinearity, by inducing a dependence of the measurement operator on the previous output state. This results in a nonlinear form of the output, i.e. $(x_{t-1}x_t)$, already after one step (for the full analytical derivation, see Supplementary Material note II), and also in a memory of the previous states. From an operational point of view, the internal variable of the memristor $R_t$, which can be seen as the reflectivity of a beam-splitter, is updated via:

\begin{equation}
    R_t=\frac{1}{m}\sum_{T=t-m}^t R_T=\frac{1}{m}\sum_{T=t-m}^t (a \cdot p_{T, 2}+b)
    \label{eq:update_P}
\end{equation}

where $p_{T, 2}$ is the probability of detecting a photon in the update mode and the internal phase of the MZI that implements $\textbf{U}_{mem}$ is given by $2 \times arccos(\sqrt{R})$. The rotations $\textbf{U}_1$ and $\textbf{U}_2$, along with $a$ and $b$, are optimizable hyperparameters (see Supplementary Material note IIA for further details). In this case, $m$ can be seen as the \textit{memory extent}, which amounts to the number of previous steps which are used when performing the average reported in Eq.~\eqref{eq:update_P}, to retrieve the reflectivity at each time step $t$.

The goal of the second task is to predict the output of a nonlinear dynamical system, 
introduced in \cite{atiya_new_2000, hochreiter_long_1997}, called \textit{NARMA}. This was already widely used as a benchmark for classical \cite{inubushi_reservoir_2017, akai-kasaya_performance_2022} and quantum \cite{fujii_harnessing_2017, fry_optimizing_2023} RC models. The recurrence relation of the system is described by the following equation:
\begin{equation}
    y_{t+1} = 0.4 y_t + 0.4 y_t y_{t-1} + 0.6x_t^3 + 0.1,
    \label{eq:narma}
\end{equation}
where $x_t$ and $y_t$ is the input and the output, respectively, at time $t$. From Eq.~\eqref{eq:narma}, it is visible that the output $y_{t+1}$ depends on past two outputs $y_{t}$ and $y_{t-1}$.  

The time series $\bm{x} = \{x_t\}_{t=1}^N$ is generated by sampling each $x_t$ independently and uniformly in the interval $U(0, \frac{1}{2})$, and then the outputs $\bm{y} = \{y_t\}_{t=1}^N$ come sequentially from Eq.~\eqref{eq:narma}.  For this task, we are using the amplitude encoding $x_t\mapsto x_t \ket{0} + \sqrt{1-x_t^2}\ket{1}$ and the feedback rule
\begin{equation}
    R_t = R_{t-1} + \frac{p_{t-1,2} - R_{t-1}}{m_d},
    \label{eq:feedback_rule}
\end{equation}
where $p_{ti}$ at time $t$ is the probability of detecting one photon in the $i$-th mode ($i=0,1,2$), while $m_d$ is the fixed \textit{memory decay} of the memristor. The memory decay, analogously to the memory extent mentioned before, is also a hyperparameter, but it is related to how fast the memory of past outcomes will decay. Indeed, this feedback rule does not involve any average and, as detailed in the Supplementary Material note IIB, it modulates the dependence of $R_t$ on each previous output.

The other two tasks are \textit{one-step-ahead} predictions of a real data series. This implies that, given $y_t$, the model needs to predict $y_{t+1}$. The first of them is the solution to the \textit{Mackey-Glass} differential equation \cite{shahi2022prediction,fujii_harnessing_2017} and the last is the so-called \textit{Santa Fe} time series, which is a dataset recorded from a far-infrared laser in a chaotic state \cite{hubner1989dimensions, weigend1993results}. For both tasks, we use amplitude encoding $x_i \mapsto \sqrt{1-x_i} \ket{0} + \sqrt{x_i} \ket{1}$ and the feedback rule in Eq. \eqref{eq:feedback_rule} with $m_d=2$ for Mackey-Glass task and $m_d=6$ for Santa Fe task. Both data sets have been scaled down to fall within the range of 0 to 1.
For more details see the Supplementary Material note IIB.

For all the tasks, after the quantum processing, a classical linear regression is used to find the best prediction. We note that during this process no nonlinearity or memory is added. As a benchmark, we performed the monomial prediction tasks without any feedback loop and with no reservoir. The latter is a fully classical procedure, only injecting the \textit{x} values into the final linear regression, to highlight the role of our physical platform. For the time series predictions, instead, we performed a comparison with slightly more complex fully classical models, even featuring some nonlinearity and memory of past states.

\begin{figure}[t]
     \centering
     \includegraphics[width=0.99\linewidth]{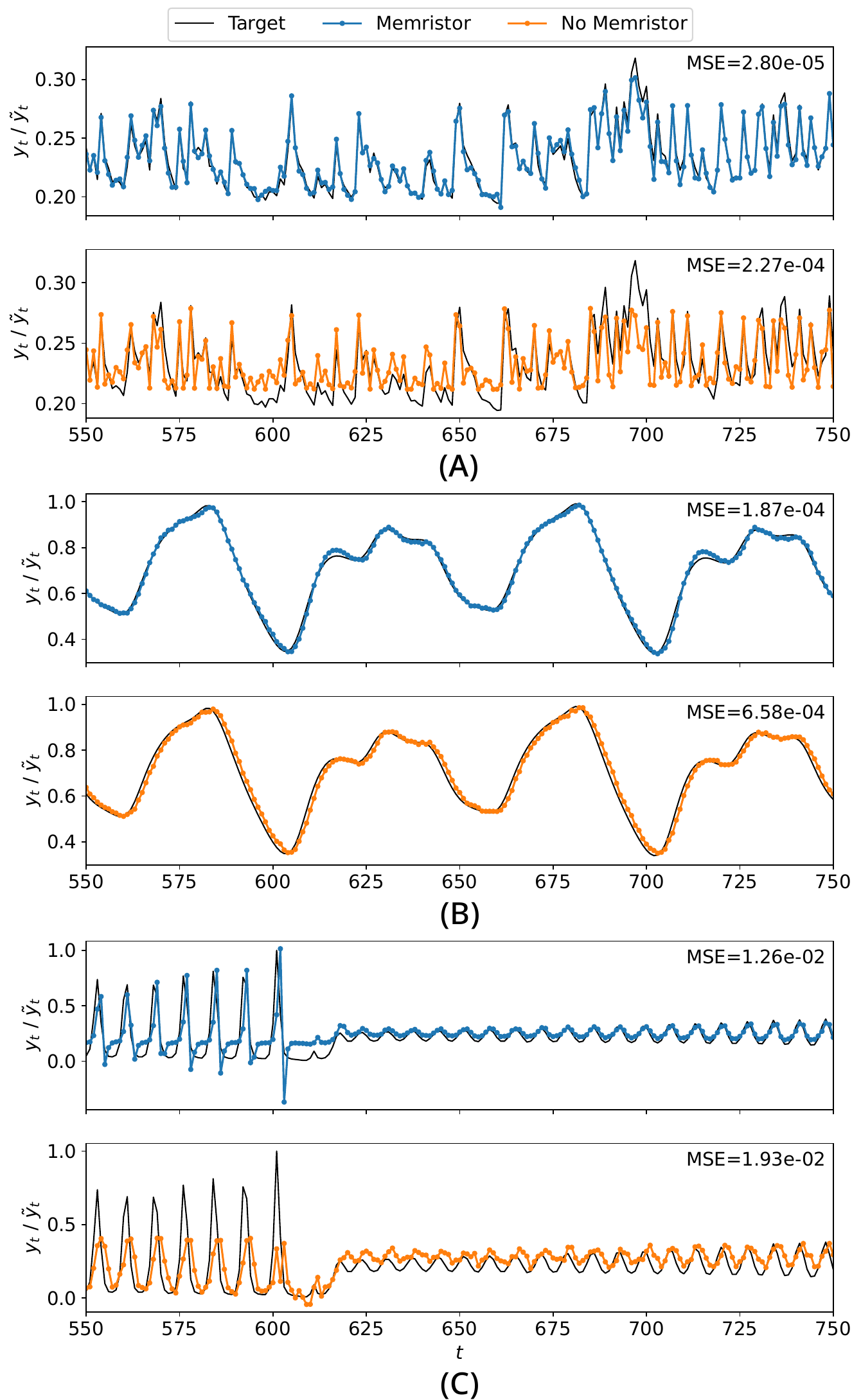}
      \caption{\textbf{NARMA, Mackey-Glass, and Santa Fe task results.}  We show the performance of our model on the following prediction tasks: NARMA (A), Mackey-Glass (B), and Santa Fe (C). The true values of the time series are plotted in black. Predictions with and without the memristor are shown, respectively, in blue and orange. In each subplot, the MSE score is reported in the top right corner. Each of the adopted datasets contains 1000 points. Out of those, the first 20 correspond to the so-called \textit{washout}, i.e. the time required to the model to initialize. Then, the training process is carried out on the following 480 points, while the validation is performed on the rest of the dataset. For the sake of clarity, the plot shows only 200 points of the test. }
     \label{fig:narma_task}
\end{figure}

\section{Experimental implementation}
The main part of the experimental setup, fully represented in Fig.~\ref{fig:experimental_setup}, is the physical quantum reservoir for information processing. It consists of an integrated waveguide circuit with 3 input/output modes and is designed according to a universal architecture \cite{clements2016optimal}. The waveguides are fabricated by direct femtosecond laser writing \cite{spagnolo_experimental_2022, gattass2008femtosecond, corrielli_femtosecond_2021} in an alumino-borosilicate glass substrate and feature single-mode operation at 1550 nm. Microstructured thermo-optic phase shifters are integrated on-chip, to provide efficient phase tunability \cite{ceccarelli_low_2020}. Let us highlight that, due to the limited size, our circuit does not suffer from thermal crosstalk (for more detailed information about the calibration procedure, see the Supplementary Material note III). Furthermore, due to the low birefringence of the substrate, this circuit is polarization independent. Input and output ports are pigtailed with single-mode fibers.

The encoding of the classical input is set via an initial unitary transformation $\textbf{U}_{enc}$, through an on-chip MZI, as shown in Fig.~\ref{fig:conceptual-scheme}b. For example, to encode the state $\sqrt{x}|0\rangle +\sqrt{1-x}|1\rangle$, apply a phase $\phi=2 \arccos(x)$ when injecting a photon into the lower mode of the MZI. However, since the optical circuit consists of three MZIs in total, we implement the product $\textbf{U}_1  \cdot\textbf{U}_{enc}$ through the first MZI, as detailed in Fig.~\ref{fig:experimental_setup}b.

The input single photon state is generated by a collinear Type-II spontaneous parametric down conversion source, pumped at 775~nm, which emits pairs of degenerate photons with a wavelength of 1550~nm. The employed nonlinear crystal is a periodically poled titanyl phosphate, placed within a Sagnac interferometer. Since for this application we do not need entangled states, the crystal is pumped only in one spatial direction and generates the separable state $|01\rangle$. One photon is ancillary and heralds the generation of the second photon that is injected into the photonic processor. This implies that we post-select our statistics, considering only the cases where both the heralding and signal photons are detected. The three output modes of the photonic processor are connected to superconducting nanowire single photon detectors. 

At this point, the frequencies of the clicks registered by the three detectors at the output of the chip, in coincidence with the one detecting the heralding photon, give the probabilities $p_0$, $p_1$ and $p_2$ as $p_i=\frac{N_{i,a}}{\sum_{j=1}^3 N_{j,a}}$.

\section{Results}

To quantify the quality of the performance of predictive tasks, we use the so-called \textit{mean squared error} \cite{mood1950introduction}, defined as follows:
\begin{equation}
    MSE(\bm{y}, \tilde{\bm{y}}) \coloneq 
    \frac{1}{N}\sum_{t=1}^N \left( y_t - \tilde{y}_t \right)^2
    \label{eq:nmse}
\end{equation}
with $y_t$ target values, $\tilde{y}_t$ predicted values and $N$ data points.

For the monomial prediction tasks three different cases are compared: (i) with quantum memristor, (ii) without memory (with no feedback loop) and (iii) with no reservoir. The prediction of $x^n$ with $n=4$ is shown in Fig.~\ref{fig:final_fp} and the inset plot shows the loss function of the readout unit for different $n$. Note that even without the feedback loop some nonlinearity can be generated, due to the encoding from $x$ to $\rho_x$. However, the feedback loop enhances the impact of the nonlinear measurement process and ultimately the performance of the algorithm. Let us highlight that, when comparing the cases with and without feedback, we are picking the best hyperparameters (namely $U_1$ and $U_3$ and the coefficients of the feedback function), separately for the two cases, to have a fair comparison.
In contrast, in the case of no reservoir, the output is simply a linear function, since the final linear regression only gives a weighted sum of the inputs.

\begin{table}[t]
    \centering
    \renewcommand{\arraystretch}{1.4} 
    \begin{tabular}{c|c|c}
        MODEL               & EQUATION & MSE ($\times 10^{-4}$) \\ \hline\hline
        $L$              & $y_{t} \approx \text{Poly}_1(x_t)$ & $2.76(29)$ \\ 
        $C$           & $y_{t} \approx \text{Poly}_3(x_t)$ & $2.05(22)$ \\ 
        $L+M$        & $y_{t} \approx \text{Poly}_1(x_t, x_{t-1})$ & $1.83(19)$ \\ 
        $C+M$     & $y_t \approx \text{Poly}_3 (x_t, x_{t-1})$ & $0.92(11)$ \\ 
        \textbf{QMEM} & - & $\bm{0.33(8)}$ \\
    \end{tabular}
    \caption{\textbf{Mean Squared Errors for classical machine learning models for NARMA task}. This table shows the average MSE achieved by several classical models, which have access to a similar amount of resources, in terms of memory and free variables, in comparison to the one presented in this work, indicated by QMEM. The numbers in the bracket indicate one standard deviation of 100 runs. $\text{Poly}_d(a_1,\ldots, a_n)$ is the polynomial of variables $\{a_1,\ldots, a_n\}$ of degree at most $d$. In detail, $L$ and $C$ stand, respectively, for models that can base their predictions on linear and cubic manipulation of the input. The first features 2 and the second 4 free variables.  $L+M$ ($C+M$) indicates the same models, but with access to the previous element of the output series. In this case, the first features 3 and the second 9 free variables. For an accurate description of such models, see Supplementary Material note II. The reported values are obtained through numerical simulations. We note that polynomials of higher order improve neither of the results. Our model features only 1 free variable, given by the memory decay $m_d$ (see Eq.~\eqref{eq:feedback_rule}). }
    \label{tab:MSE_models}
\end{table}

 In Fig. \ref{fig:narma_task}, we show the experimental prediction of the network for the previously described time series prediction tasks. 
 For the case with no feedback loop, each output depends only on the last input, since there is no storing the information of previous data points. 
It is worth noting that cases without feedback loops still achieve fairly good results (although worse than the case where the quantum memristor is employed). For the NARMA task, this is due to its low nonlinearity, while for the Mackey-Glass and Santa Fe tasks, it results from the high density of probing points, which makes the prediction almost linear.
To have a more complete overview of the performance of our algorithm, we present the results achieved on the same tasks by simple classical algorithms. First, we consider a model without memory, where the prediction is a polynomial function of the input data. In particular, we consider two models, featuring 2 and 4 free parameters. Next, we examine two models with the simplest possible memory, which records only the previous input. Hence, the final prediction will be given by a polynomial function of the current and previous input. In this case, the two models have, respectively, 3 and 9 free parameters. Our model featuring the quantum memristor has a slightly better performance than those models (see Table \ref{tab:MSE_models}), even when the classical one has access to past inputs. For the comparison of the other tasks and the mathematical description of these models, see Supplementary Material note IIC.

\section{Discussion}
In this work, we have implemented the first instance of QRC exploiting single photon states based on a quantum memristor device. These machine learning models work by injecting input data into a fixed random quantum reservoir, implemented through a photonic quantum memristor, which is then followed by a classical linear regression model \cite{mujal_opportunities_2021}. 
 We tested our algorithm on \changes{four} function prediction tasks: several monomial functions, and then NARMA, Santa Fe and Mackey-Glass time series. Due to its regularity, the first case is only to test the nonlinearity that is achievable through our apparatus. The others, instead, test also the memory of our model, given that its outcomes depend on previous ones.

Ours is the first example of neuromorphic architectures exploiting single-photon states (encoded in the path degree of freedom) and implementing a nonlinear behaviour through an adaptive protocol, performed by the quantum memristor. Moreover, we carry out a systematic analysis to distinguish the source of such a nonlinearity, i.e. the encoding or the feedback loop. For this reason, for all of the tasks, we compare the case featuring the quantum memristor (QRC), to the case without feedback loop, consistently showing that the quantum memristor brings an enhanced accuracy with respect to classical models. This scheme, although applied to relatively simple tasks, can be seen as a building block for more complex networks, as the output of the memristor, when tracing out the update mode, preserves coherence and can be used as an input for successive layers (see the Supplementary Material note I for further details). 

A potential way of enhancing the expressivity of the proposed model is to use multiple memristors in a consecutive way, as shown for instance in \cite{di2025quantum}, and updating both the internal variables of those memristors, as well as the input state, so that new inputs would interfere with states encoding previous outcomes. Analogously, the implementation of a fast feedback would allow to make coherent adjustments to the quantum state. This would result in a larger memory storage and open up to the possibility of tackling a wider range of tasks. Moreover, feeding parts of the outputs as new inputs, for instance through a loop architecture \cite{carosini2024programmable}, could unlock the possibility of tackling forecasting tasks \cite{makridakis2018statistical}.  Another interesting feature is that a quantum reservoir could be used to directly process quantum systems, tackling tasks with no classical equivalent. For instance, to investigate foundational aspects or quantum state properties \cite{suprano2024experimental, cong2019quantum, cimini2021calibration}.
 
Besides exploring the possibilities unlocked by the quantum features of the memristor, this study paves the way for further investigations related to the nonlinearities achievable on photonic platforms, for optical computing. This is particularly relevant, considering that hybrid optical-electronic ANNs have been proven to require a lower energy consumption than standard ones \cite{mcmahon_physics_2023, hamerly2019large}. In this framework, the memristor could be employed to constitute the activation function layer of quantum neural networks and, after further investigations on the relation between the memristor's dynamics with neurons one, it could unlock the possibility of implementing spiking optical neural networks \cite{brand2024quantum, rungratsameetaweemana2025random, yamazaki2022spiking}.

\section*{Acknowledgments}
The fabrication of the photonic integrated circuit was partially carried out at PoliFAB (https://www.polifab.polimi.it/), the micro- and nano-fabrication facility of Politecnico di Milano. F.C. and R.O. would like to thank the PoliFAB staff for technical advice and support. This research was funded in whole or in part by the Austrian Science Fund (FWF)[10.55776/ESP205] (PREQUrSOR), [10.55776/F71] (BeyondC), [10.55776/FG5] (Research Group 5) and [10.55776/I6002] (PhoMemtor).
R.O. acknowledges financial support by ICSC – National Research Center in High Performance Computing, Big Data and Quantum Computing, funded by European Union– NextGenerationEU.

\section*{Author Contribution}
Mir.S. and I.A. designed and conducted the experiment, Mich.S. developed the theory and algorithm for the time series predictions and analyzed the corresponding data, I.A. developed the algorithm for power predictions and I.A. and Mir.S. analyzed the corresponding data. R.A., A.C. and F.C. conducted the design, fabrication and calibration of the integrated photonic processor. Mir.S., I.A. and Mich.S. wrote the first draft of the manuscript. 
I.A., Mag.S., R.O., B.D. and P.W. supervised the whole project.
All authors discussed the results and reviewed the manuscript.

\section*{Competing interests}
F.C. and R.O. are co-founders of the company Ephos. The authors declare that they have no other competing interests.


\end{document}


\title{Supplementary Material} 

\maketitle

\section{On-chip quantum tomography of quantum memristor output state}

As detailed in \cite{spagnolo_experimental_2022} and in the main text, the photonic quantum memristor has two features: (i) \textit{memristive behaviour} and (ii) \textit{quantum coherent processing}. These characteristics seem in contradiction, as the evolution of closed quantum systems is intrinsically linear, while open systems have low coherence.
However, the weak measurement scheme implemented by our device allows to express nonlinear behavior, while preserving some coherence.

\begin{figure*}[tb]
    \centering
    \hspace*{\fill}
    \begin{minipage}[c]{0.27\textwidth}
        \centering
        \includegraphics[width=\textwidth]{  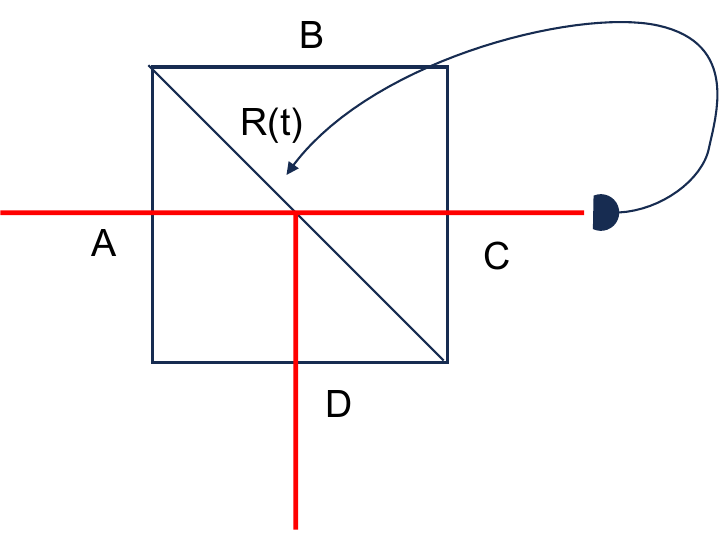}
        (a)
        \label{fig:ann-activation-f}
    \end{minipage}
    \hspace*{\fill}
    \begin{minipage}[c]{0.63\textwidth}
        \centering
        \includegraphics[width=\textwidth]{  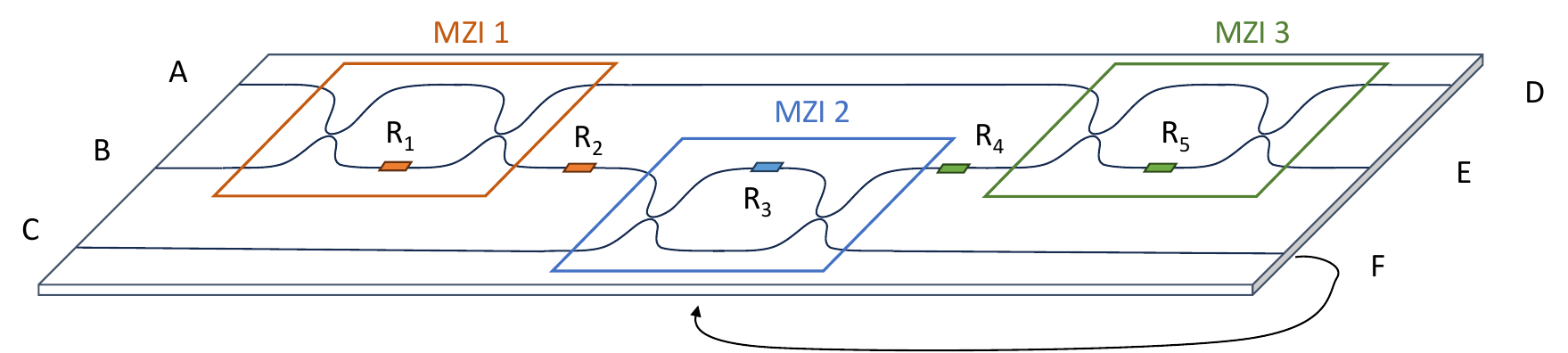}
        (b)
        \label{fig:mbs}
    \end{minipage}
    \caption{\textbf{Quantum memristor schemes}. (a) \textbf{Original scheme for the quantum memristor.} This device is seen as a beam-splitter with a variable reflectivity, which is updated depending on the measurements performed on port C. In this configuration, the input state is a superposition of vacuum and a single photon state, injected into mode A, i.e. $\cos(\alpha) |0\rangle_A+ \sin(\alpha)|1\rangle_A$. (b) \textbf{Quantum memristor integrated circuit.} In this case, the variable beam-splitter is replaced by a Mach-Zehnder interferometer (MZ2) and the input state is now encoded in the dual-rail encoding, as follows: $\cos(\alpha) |0\rangle_{AB}+ \sin(\alpha)|1\rangle_{AB}$. This state, in the Fock basis, reads as $\cos(\alpha) |10\rangle_{AB}+ \sin(\alpha)|01\rangle_{AB}$.}
    \label{fig:SM-memristor}
\end{figure*}

Let us consider the original scheme of the photonic quantum memristor (see Fig.~\ref{fig:SM-memristor}a), namely a beam splitter with variable reflectivity, where we inject a coherent superposition of vacuum and a single photon state into input mode A, i.e. $cos (\alpha) |0\rangle_A + e^{(i\phi)} sin(\alpha)|1\rangle_A$. Then, one of the two output modes (C), is measured and the reflectivity of the beam splitter is varied according to a feedback loop.

In our case, we simulate this input state, using the circuit in Fig.~\ref{fig:SM-memristor}b, where we use the first two spatial modes (A and B) to encode a qubit state, with dual rail encoding, i.e. $cos (\alpha) |0\rangle + e^{(i\phi)} sin(\alpha)|1\rangle$. For convenience, we will express this state in terms of occupational numbers, i.e.  $cos (\alpha) |10\rangle_{AB} + e^{(i\phi)} sin(\alpha)|01\rangle_{AB}$. 
Then, we place a variable beam splitter, i.e. a Mach-Zehnder interferometer, acting between modes $B$ and $C$ (MZI2). The state will evolve then as follows:  
\begin{equation}
\begin{split}
    \rho_{out} = &|cos(\alpha)|^2 |100\rangle_{DEF}\langle100| + 
    cos(\alpha)sin(\alpha)e^{-i\phi}\sqrt{1-R}|100\rangle_{DEF}\langle010|-\\
    &icos(\alpha)sin(\alpha)e^{-i\phi}\sqrt{R}|100\rangle_{DEF} \langle 001|+\\
    &cos(\alpha)sin(\alpha)\sqrt{1-R}|010\rangle_{DEF}\langle100|+|sin(\alpha)|^2(1-R)|010\rangle_{DEF}\langle 010|-\\
    &i|sin(\alpha)|^2\sqrt{R(1-R)}|010\rangle_{DEF}\langle 001|+\\
    &icos(\alpha)sin(\alpha)\sqrt{R}|001\rangle_{DEF}\langle100|+
    i|sin(\alpha)|^2\sqrt{R(1-R)}|001\rangle_{DEF}\langle010|+\\
    &|sin(\alpha)|^2R|001\rangle_{DEF}\langle001|
\end{split}
\end{equation}

where $R$ is the reflectivity of the beam splitter (corresponding to $cos^2(\frac{\theta}{2})$, if we consider the internal phase of MZI2).

At this point, we want to quantify the coherence of the output state of the memristor, tracing out the part used for the feedback loop. Namely, we are interested in the state coming out of the first two modes of the circuit in Fig.~\ref{fig:SM-memristor}b, D and E, or, analogously from port D of Fig.~\ref{fig:SM-memristor}a.
To do this, we trace out mode F and obtain the following state:

\begin{equation}
\begin{split}
    \rho_{out} = &|cos(\alpha)|^2 |10\rangle_{DE}\langle10| +\\ 
    &|sin(\alpha)|^2(1-R)|01\rangle_{DE}\langle 01|+\\
    &|sin(\alpha)|^2R|00\rangle_{DE}\langle00|+\\
    &cos(\alpha)sin(\alpha)e^{-i\phi}\sqrt{1-R}|10\rangle_{DE}\langle01|+\\
    &cos(\alpha)sin(\alpha)\sqrt{1-R}|01\rangle_{DE}\langle10|
\end{split}
\end{equation}

To be consistent with the original scheme of the quantum memristor, we must note that the two states $|10\rangle_{DE}$ and $|00\rangle_{DE}$ both correspond to the state $|0\rangle_D$ of the original scheme (see, respectively, Fig.~\ref{fig:SM-memristor}a and  Fig.~\ref{fig:SM-memristor}b). Indeed, both states correspond to vacuum at the non-measured output of MZI2. Hence, the density matrix is the following:

\begin{equation}
\begin{split}
    \rho^*_{out} = &|cos(\alpha)|^2 |0\rangle\langle0| +\\ 
    &|sin(\alpha)|^2(1-R)|1\rangle\langle 1|+\\
    &|sin(\alpha)|^2R|0\rangle\langle0|+\\
    &cos(\alpha)sin(\alpha)e^{-i\phi}\sqrt{1-R}|0\rangle\langle1|+\\
    &cos(\alpha)sin(\alpha)\sqrt{1-R}|1\rangle\langle0|
\end{split}
\end{equation}

The purity of this state is then:

\begin{equation}
 Tr(\rho^{*2}_{out}) = 1-2sin(\alpha)^4R(1-R)
 \label{eq:purity_mem}
\end{equation}

\begin{figure}[tb]
        \centering
        \includegraphics[width=\linewidth]{  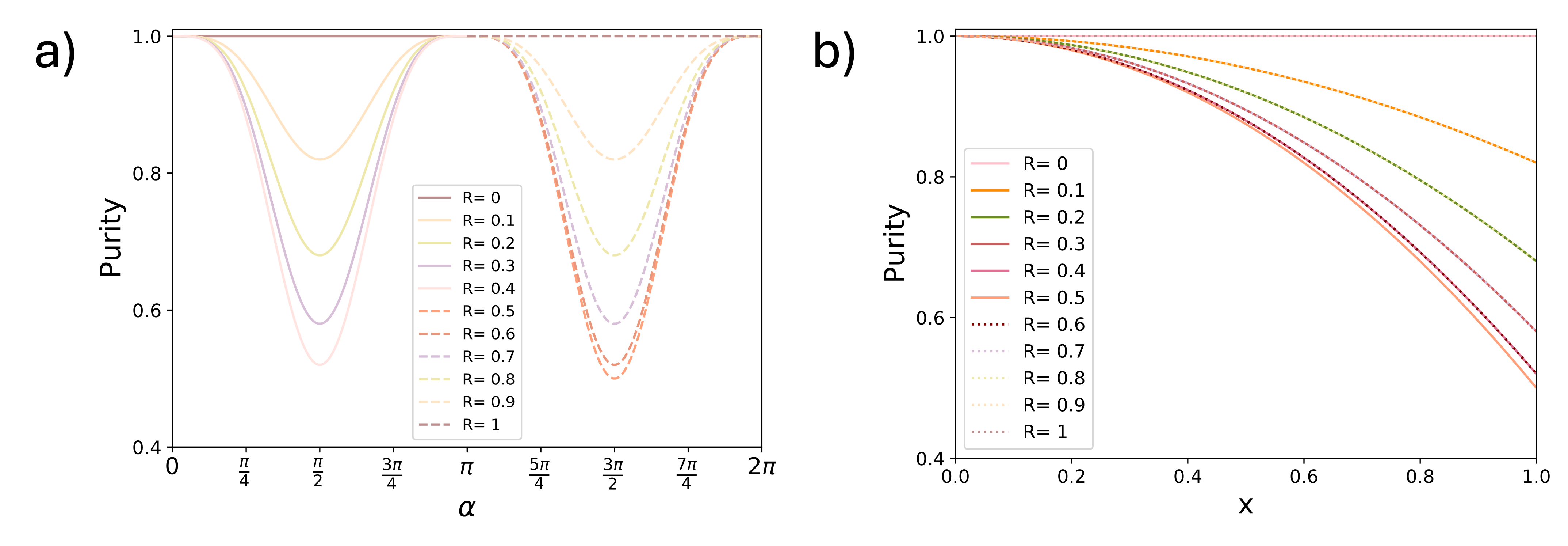}
        \caption{\textbf{Partial output state purity}. (a) Purity dependence on the $\alpha$ parameter, considering the input state as $cos(\alpha)|10\rangle_{AB}+ sin(\alpha)|01\rangle_{AB}$. The behavior is shown for different values of $R$ and, as expected, it is periodic with an interval of $\pi$. For the sake of clarity, for $R\geq0.5$, the behaviour is shown in the interval from $\pi$ to $2\pi$ with dashed curves. Note that, for reflectivities $R=r$ and $R=1-r$, e.g. $R=0.1$ and $R=0.9$, the curves overlap. The lowest purity is obtained for $R=0.5$ and $|sin(\alpha)|=1$ (i.e. $\alpha=\frac{\pi}{2}+i\times \pi$, for $i=0,1,...$), as it can be derived from Eq.~\eqref{eq:purity_mem}. (b) We show the same behaviour as in the previous plot, but considering our classical data encoding, i.e. $\sqrt{1-x}|1,0\rangle_{AB}+ \sqrt{x}|0,1\rangle_{AB}$. For values $R\leq0.5$ we use solid lines and, for $R>0.5$, dashed lines, to show the overlap for reflectivities $R=r$ and $R=1-r$, e.g. $R=0.1$ and $R=0.9$.   Also in this case, it is visible that the lowest purity is displayed for $|x|=1$ and $R=0.5$ (see Eq.~\eqref{eq:purity_exp}).}
        \label{fig:purity}
    \end{figure}


From Eq.~\eqref{eq:purity_mem}, we note that the output state is pure, when the reflectivity is either 0 or 1. Indeed in the extremal case when $R=0$, the input remains unchanged, as the memristor performs the identity transformation. Instead, when $R=1$, the portion of the state that goes in the memristor is completely redirected to the measured mode F (or C). Hence, when tracing it out and looking only at the other output mode of the beam splitter/MZI, the only detected state is vacuum. The expected values of the purity for different reflectivities and input states (i.e. $\alpha$) are shown in Fig.~\ref{fig:purity}. Let us note that, due to the symmetric form of Eq.~\eqref{eq:purity_mem}, the curves for reflectivities $R$ and $1-R$ overlap.

\begin{table}[b]
    \centering
    \begin{minipage}{0.3\textwidth}
        \centering
        \begin{tabular}{c|c|c|c}
            $x=0.1$ &  $R=0$ & $R=0.5$ & $R=1$\\ \hline
       $Tr(\rho^{*2}_{out\ exp})$ & 1.00 & 0.95 & 0.99 \\
       $Tr(\rho^{*2}_{out\ th})$  & 1.00 & 0.99 & 1.00 \\
        \end{tabular}
    \end{minipage}%
    \hspace{0.05\textwidth} 
    \begin{minipage}{0.3\textwidth}
        \centering
        \begin{tabular}{c|c|c|c}
            $x=0.5$ &  $R=0$ & $R=0.5$ & $R=1$\\ \hline
       $Tr(\rho^{*2}_{out\ exp})$ & 0.96 & 0.82 & 0.98 \\
       $Tr(\rho^{*2}_{out\ th})$  & 1.00 & 0.87 & 1.00 \\
        \end{tabular}
    \end{minipage}%
    \hspace{0.05\textwidth} 
    \begin{minipage}{0.3\textwidth}
        \centering
        \begin{tabular}{c|c|c|c}
            $x=0.9$ &  $R=0$ & $R=0.5$ & $R=1$\\ \hline
       $Tr(\rho^{*2}_{out\ exp})$ & 0.93 & 0.61 & 0.99 \\
       $Tr(\rho^{*2}_{out\ th})$  & 1.00 & 0.59 & 1.00 \\
        \end{tabular}
    \end{minipage}
    \caption{\textbf{Experimental purity on partial output state.} We show the values obtained experimentally, from the reconstructed density matrix of the output state $\rho_{DE}$, when tracing out the mode used for the feedback loop (F). These values are then compared to the expected ones, obtained through Eq.~\eqref{eq:purity_exp}.}
    \label{tab:purityvalues}
\end{table}

Experimentally, we evaluated the purity for three different input states $\sqrt{1-x}|0\rangle +\sqrt{x}|1\rangle$ (with $x=0.1, 0.5, 0.9$) and three values of $R$.
To this aim, we performed a quantum state tomography on the output state in modes $D$ and $E$ of the circuit, exploiting MZI1 for the encoding MZI3 for the measurements. MZI2 was used to select the reflectivity (see Fig.~\ref{fig:SM-memristor}b).

From a practical point of view, to produce each state $\sqrt{1-x}|0\rangle +\sqrt{x}|1\rangle$ and reflectivity $R$, we injected a single photon from mode B of our circuit and set the internal phase of MZI1 to $2\times arccos (\sqrt{1-x})$ and MZI2 as $2\times arccos (\sqrt{R})$.
Then, to perform the quantum state tomography, we implemented projective measurements onto the Pauli operators bases, i.e. $\sigma_x$, $\sigma_y$ and $\sigma_z$.
The corresponding internal/external phases ($\phi$/$\psi$) for MZI3 were, respectively, ($0$/$\frac{\pi}{2}$), for $\sigma_x$, ($\frac{\pi}{2}$/$\frac{\pi}{2}$), for $\sigma_y$ and ($0$/$0$), for $\sigma_z$.

\begin{figure}[!h]
    \centering
    \includegraphics[width=0.8\linewidth]{  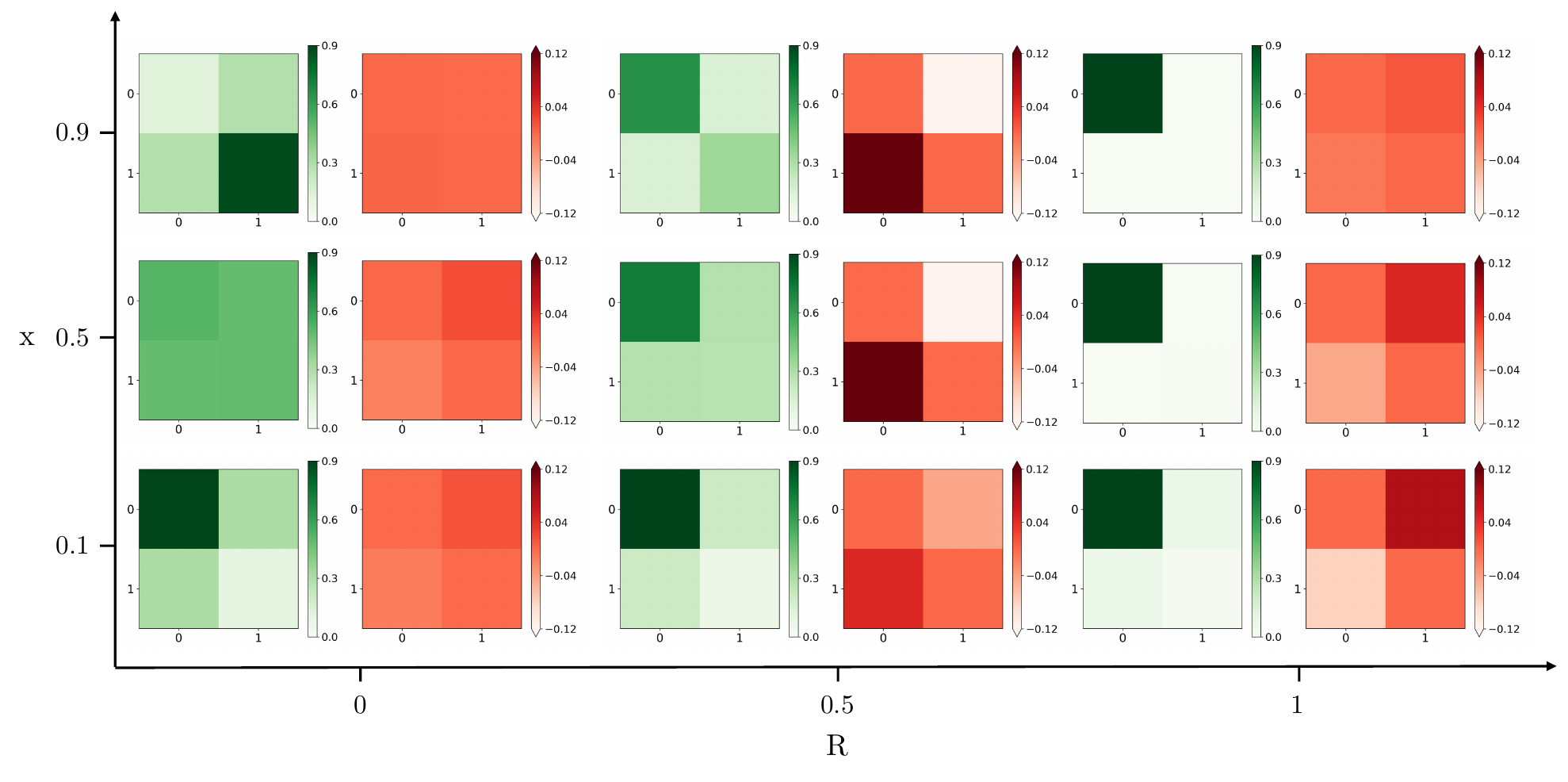}
    \caption{\textbf{Experimental density matrix of the memristor partial output state.} We performed the quantum state tomography on the state exiting from the first two modes of our circuit (D and E), for several values of the reflectivity of the memristor, i.e. $R=0, 0.5, 1$, and input states $\sqrt{1-x}|1,0\rangle_{AB}+\sqrt{x}|0,1\rangle_{AB}$, i.e. $x=0.1, 0.5, 0.9$. In green, we show the colourplot of the real part of the experimental density matrix, while, in orange, the imaginary part. Let us note that, on the element $|0\rangle\langle0|$ we manually added the term $x_{exp}^2 R=p_F$, to be consistent with the original model of the quantum memristor.}
    \label{fig:tomo_exp}
\end{figure}

\begin{figure}[!h]
    \centering
    \includegraphics[width=0.8\linewidth]{  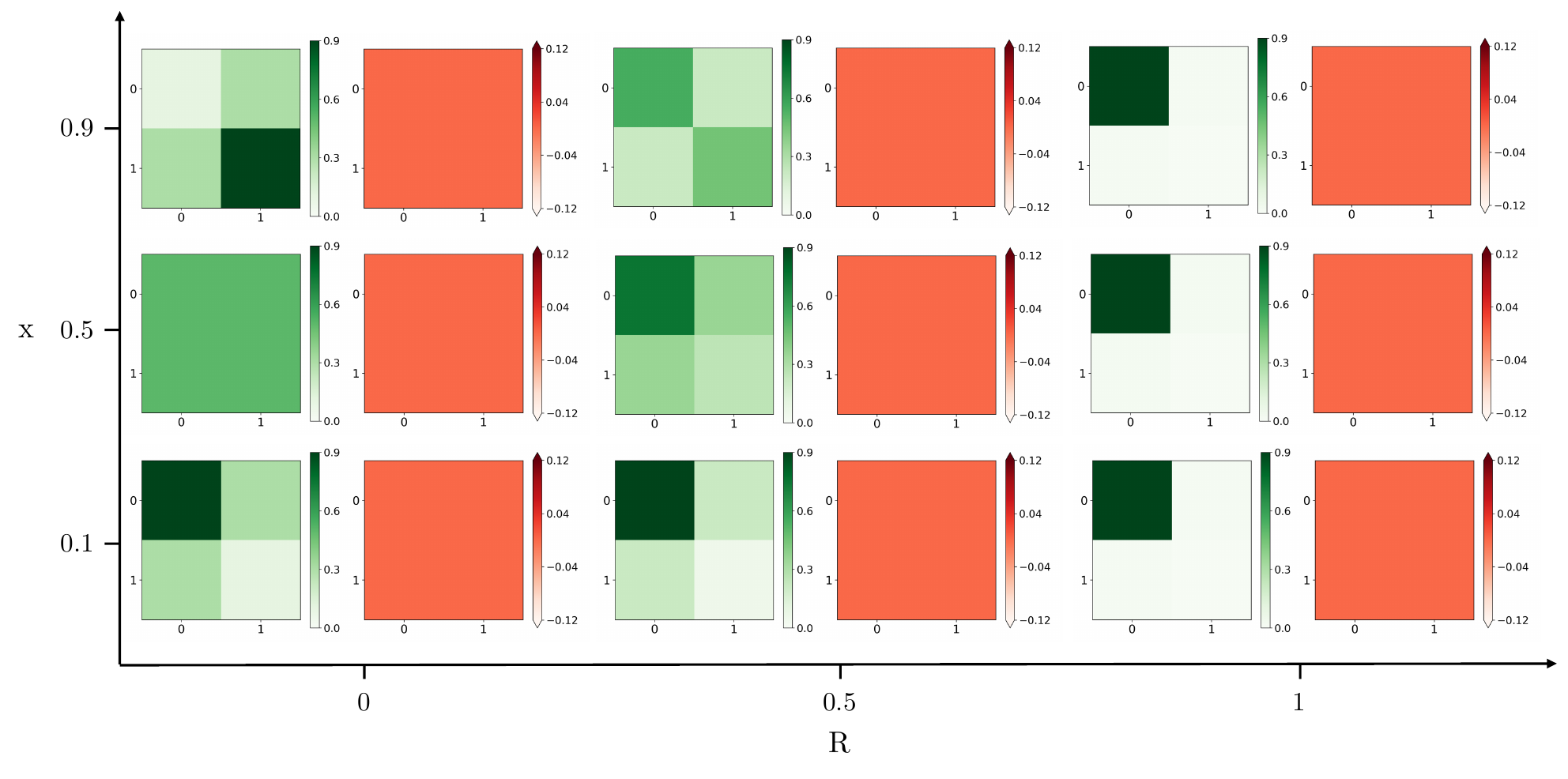}
    \caption{\textbf{Expected density matrix of the memristor partial output state.} We consider the memristor output state, tracing out mode F, for several values of the reflectivity of the memristor, i.e. $R=0, 0.5, 1$, and input states $\sqrt{1-x}|1,0\rangle_{AB}+\sqrt{x}|0,1\rangle_{AB}$, i.e. $x=0.1, 0.5, 0.9$. In green, we show the colourplot of the real part of its density matrix, while, in orange, the imaginary part.}
    \label{fig:tomo_th}
\end{figure}

The resulting experimental density matrices are shown in Fig.~\ref{fig:tomo_exp}. The expected ones ($\rho_{exp}$), instead, are shown in Fig.~\ref{fig:tomo_th}. In the Fock notation, $\rho_{exp}$ amounts to: 

\begin{equation}
\begin{split}
    \rho_{exp} = &x |10\rangle\langle10| +\\ 
    &(1-x)(1-R)|01\rangle\langle 01|+\\
    &\sqrt{x(1-x)}e^{-i\phi}\sqrt{1-R}|10\rangle\langle01|+\\
    &\sqrt{x(1-x)}\sqrt{1-R}|01\rangle\langle10|
\end{split}
\end{equation}

Let us note that, since we post-select events where we detect one photon, either on output mode D or E, the term $x^2R|00\rangle\langle00|$ is not directly measured in the tomography. Hence, to retrieve the purity, we evaluated the experimental value of $x$ ($x_{exp}$) from the density matrix and then applied to Eq.~\eqref{eq:purity_mem}, which now reads as:

\begin{equation}
 Tr(\rho^{*2}_{out}) = 1-2x_{exp}^2R(1-R)
 \label{eq:purity_exp}
\end{equation}

The experimental purity values, together with the expected ones, are reported in Table \ref{tab:purityvalues}.

\section{Photonic memristor-based Quantum Reservoir Computing}

Reservoir computing is a supervised machine learning model that, given $N_{tr}$ data points belonging to a training set $\{(x_k^{tr}, y_k^{tr})\}_{k=1}^{N_{tr}} \subset \mathbb{R}^n \times \mathbb{R}^m$, is supposed to find a target function $f_{t}: \mathbb{R}^n \rightarrow \mathbb{R}^m$, such that, $f_{t} (x_k^{train}) \simeq y_k^{train}$ for every $k$. Then, if the learning process is effective, the model should be able to generalize its operation to previously unseen data points $\{(x_k^{test}, y_k^{test})\}_k$, so that $f_{t} (x_k^{test}) \simeq y_k^{test}$. In general, this target function will not be linear and, in the reservoir computing case, it is given by the composition of two functions: $f(x)=W(g(x))$, where $W$ is a linear functional, which acts on a nonlinear function of the input $x$. The nonlinear function is implemented by the dynamics of the reservoir (also equipped with memory). Then, the parameters of $W$ are trained to minimize the distance between $W(g(x_k^{tr}))$ and $y_k^{tr}$. This last part of the algorithm amounts to a single layered artificial neural network, namely a linear regression model.

In our work, we propose a quantum reservoir scheme, where the reservoir consists in a physical system, evolved through a quantum memristor. Then, this is followed by a classical linear regression algorithm, which outputs a weighted sum of the probabilities generated by the physical system.

In order to investigate the feasibility and potentialities of the model, it is essential to study which kind of nonlinearity we can achieve with a quantum memristor device.
To this goal, let us consider a generic quantum state $\rho$, underlying a \textit{completely positive trace preserving} (CPTP) map $\Lambda$. The evolved state is then measured through a $\Sigma$-output \textit{positive operator-valued measure} (POVM), whose elements are indicate as $E_m$. In this context, the following mapping occurs:
\begin{equation}
    \rho \rightarrow p_m=Tr(E_m\Lambda(\rho))
    \label{eq:linear_equation_01}
\end{equation}
and for $m \in (1, ..., \Sigma)$, the output probability $p_m$ is always linear in the elements of the density matrix of $\rho$ \cite{innocenti_potential_2023}.
However, by adding a feedback loop, the operation applied to the state $\rho$ at time $t_n$, is affected (at least) by the output probability registered at time $t_{n-1}$. Indeed, in this case, the channel applied to the input state becomes a sequence of $\Lambda(t_n)= \Lambda(\rho(t_1), ..., \rho(t_{n-1}))$, which depends on the states that were previously injected. 

This implies then that, at time $t_n$, we will have:
\begin{equation}
    \rho \rightarrow p_m(t_n)=Tr(E_m \Lambda(\rho(t_1), ..., \rho(t_{n-1}))(\rho(t_n))) \propto \rho(t_1)\cdot \rho(t_2) \cdot... \cdot \rho(t_{n-1})
    \label{eq:linear_equation_02}
\end{equation}
This behaviour creates also an internal memory of previous states and such key features, nonlinearity and memory, give the photonic quantum memristor the necessary ingredients for a reservoir computing model.

To properly see the effect of the feedback loop on the overall achievable nonlinearity and separate it from the nonlinearity coming from the encoding, let us consider the simplest configuration of our physical system.
This consists in a single photon state, evolved through a tunable integrated circuit, whose operation is updated depending on the measurements performed on one of the output modes.
For clarity purposes, we report the scheme of the circuit below (Fig.~2b of main text), where the upper part represents the waveguides and thermal heaters of the optical circuit and the lower one describes the unitary operations that are implemented by each Mach-Zehnder interferometer.

Now, let us consider $U_1=U_3=\mathbb{I}$ and the following feedback rule: $R_{t+1}=p_{2t}$ (which is a simplified version of those used in our work).

\begin{figure*}[!h]
        \centering
        \includegraphics[scale=1.2]{  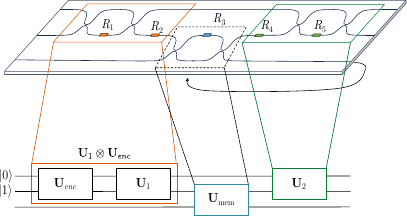}
        \label{fig:right_top}
\end{figure*}

If we consider the first step of the protocol and initialize the reflectivity of the memristor to $R_0$, the operation of the circuit (in the Fock basis, $[1,0,0]$, $[0,1,0]$ and $[0,0,1]$) will be the following:

    \[
\begin{bmatrix}
1 & 0 & 0 \\
0 & ie^{\frac{i\phi*}{2}}sin(\frac{\phi^*}{2}) & ie^{\frac{i\phi*}{2}}cos(\frac{\phi^*}{2})  \\
0 & ie^{\frac{i\phi*}{2}}cos(\frac{\phi^*}{2}) & -ie^{\frac{i\phi*}{2}}sin(\frac{\phi^*}{2}) 
\end{bmatrix}
\]

where $\sqrt{R_0}=ie^{\frac{i\phi*}{2}}cos(\frac{\phi^*}{2})$. On the other hand, the input state, at time $t=0$, will be:

\[
\begin{bmatrix}
\sqrt{x_0} \\
\sqrt{1-x_0} \\
0
\end{bmatrix}
\]

After evolving the input state through the circuit, the output probability $p_{mt}$, where $m$ is the output mode and $t$ is the time step, will be the following: $p_{00}=x_0$, $p_{10}=sin^2(\frac{\phi^*}{2})(1-x_0)$ and $p_{20}=cos^2(\frac{\phi^*}{2})(1-x_0)$.
At this point, in the \textbf{feedback} case, the operation of the circuit, at $t=1$, will be updated as follows:

    \[
\begin{bmatrix}
1 & 0 & 0 \\
0 & ie^{\frac{i\phi'}{2}}\sqrt{1-R_1} & ie^{\frac{i\phi'}{2}}\sqrt{R_1}  \\
0 & ie^{\frac{i\phi'}{2}}\sqrt{R_1} & -ie^{\frac{i\phi'}{2}}\sqrt{1-R_1} 
\end{bmatrix}
\]

where $R_1=p_{20}=cos^2(\frac{\phi^*}{2})(1-x_0)$.

\vspace{1em} 

Now, considering that the input at time $t=1$ will be $[\sqrt{x_1}, \sqrt{1-x_1}, 0]$, the output probability will be $p_{01}=x_1$, $p_{11}=sin^2(\frac{\phi^*}{2})(1-x_0)(1-x_1)$ and $p_{21}=cos^2(\frac{\phi^*}{2})(1-x_0)(1-x_1)$.

On the other hand, in the \textbf{non-feedback} case, at time $t=1$, the operation of the circuit will not be updated. Hence, the output probability at the second step will amount to $p_{00}=x_1$, $p_{10}=sin^2(\frac{\phi^*}{2})(1-x_1)$ and $p_{20}=cos^2(\frac{\phi^*}{2})(1-x_1)$. If we compare the two cases, it is evident how, even after one single step, the feedback produces a nonlinear output, that retains information about the previous outcome(s).

If we slightly complicate the scenario and we choose $U_1\neq \mathbb{I}$ and $U_3 \neq \mathbb{I}$, the evolved input state, after the time step $t=0$, will be the following: 

    \[
\begin{bmatrix}
e^{i\frac{\phi_3+\phi_1}{2}}[(sin(\frac{\phi_3}{2})sin(\frac{\phi_1}{2})+cos(\frac{\phi_3}{2})cos(\frac{\phi_1}{2})e^{i\frac{\phi_2}{2}}cos(\frac{\phi_2}{2})) \sqrt{x_0} +(-sin(\frac{\phi_3}{2})cos(\frac{\phi_1}{2})+cos(\frac{\phi_3}{2})sin(\frac{\phi_1}{2})e^{i\frac{\phi_2}{2}}sin(\frac{\phi_2}{2})\sqrt{1-x_0}]\\

e^{i\frac{\phi_3+\phi_1}{2}}[(-sin(\frac{\phi_1}{2})cos(\frac{\phi_3}{2})+sin(\frac{\phi_3}{2})cos(\frac{\phi_1}{2})e^{i\frac{\phi_2}{2}}sin(\frac{\phi_2}{2}))\sqrt{x_0} +(cos(\frac{\phi_3}{2})cos(\frac{\phi_1}{2})+sin(\frac{\phi_3}{2})sin(\frac{\phi_1}{2})e^{i\frac{\phi_2}{2}}sin(\frac{\phi_2}{2})\sqrt{1-x_0}]\\

e^{i\frac{\phi_2+\phi_1}{2}}[(cos(\frac{\phi_2}{2})cos(\frac{\phi_1}{2})\sqrt{x_0}+cos(\frac{\phi_2}{2})sin(\frac{\phi_1}{2})\sqrt{1-x_0}]\\
\end{bmatrix}
\]

where $\phi_1$, $\phi_2$ and $\phi_3$ are the internal phases of the Mach-Zehnder interferometers implementing $U_1$, $U_2$ (the memristor unitary) and $U_3$. Hence, in this case, if we measure, there will be some nonlinearity, due to the encoding, since each probability will display a term of the form $\sqrt{x_0(1-x_0)}$. However, for the \textbf{non-feedback} case, the outputs will have the same form for each time $t$, while, for the  \textbf{feedback} case, the reflectivity/transmittivity of the memristor Mach-Zehnder, i.e. $cos(\frac{\phi_2}{2})$/$sin(\frac{\phi_2}{2})$, will be updated to $e^{i\frac{\phi_2+\phi_1}{2}}[cos(\frac{\phi_2}{2})cos(\frac{\phi_1}{2})\sqrt{x_{t-1}}+cos(\frac{\phi_2}{2})sin(\frac{\phi_1}{2})\sqrt{1-x_{t-1}}]$, making the transformation more complex and more nonlinear (since it will always feature, at least, a product between $\sqrt{x_t}\sqrt{1-x_{t}}$ and $\sqrt{x_{t-1}}\sqrt{1-x_{t-1}}$). This also implies that a memory of the previous outcomes will be maintained.

\begin{figure}[tb] 
    \centering
    
    \begin{minipage}{\textwidth}
        \centering
        \includegraphics[width=0.5\textwidth]{  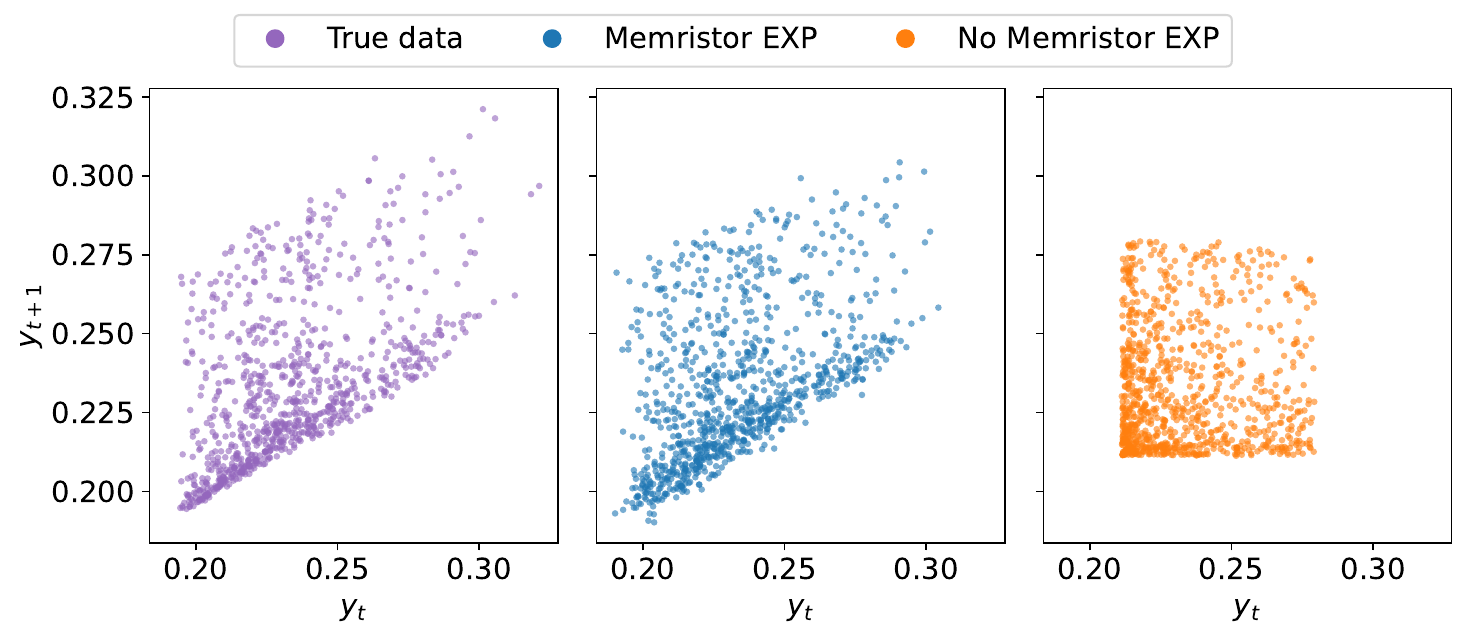}
        \vspace{1mm}
        (a)
    \end{minipage}

    \vspace{3mm}

    \begin{minipage}{\textwidth}
        \centering
        \includegraphics[width=0.5\textwidth]{  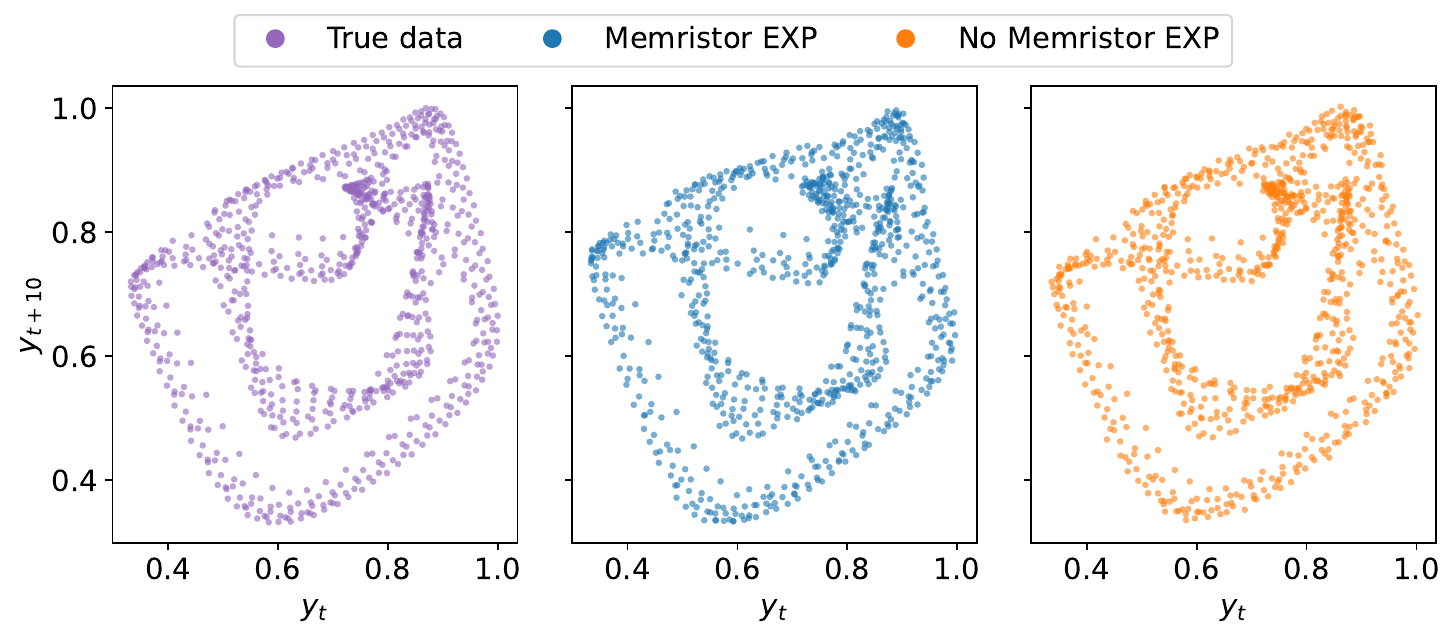}
        \vspace{1mm}
        (b)
    \end{minipage}

    \vspace{3mm}

    \begin{minipage}{\textwidth}
        \centering
        \includegraphics[width=0.5\textwidth]{  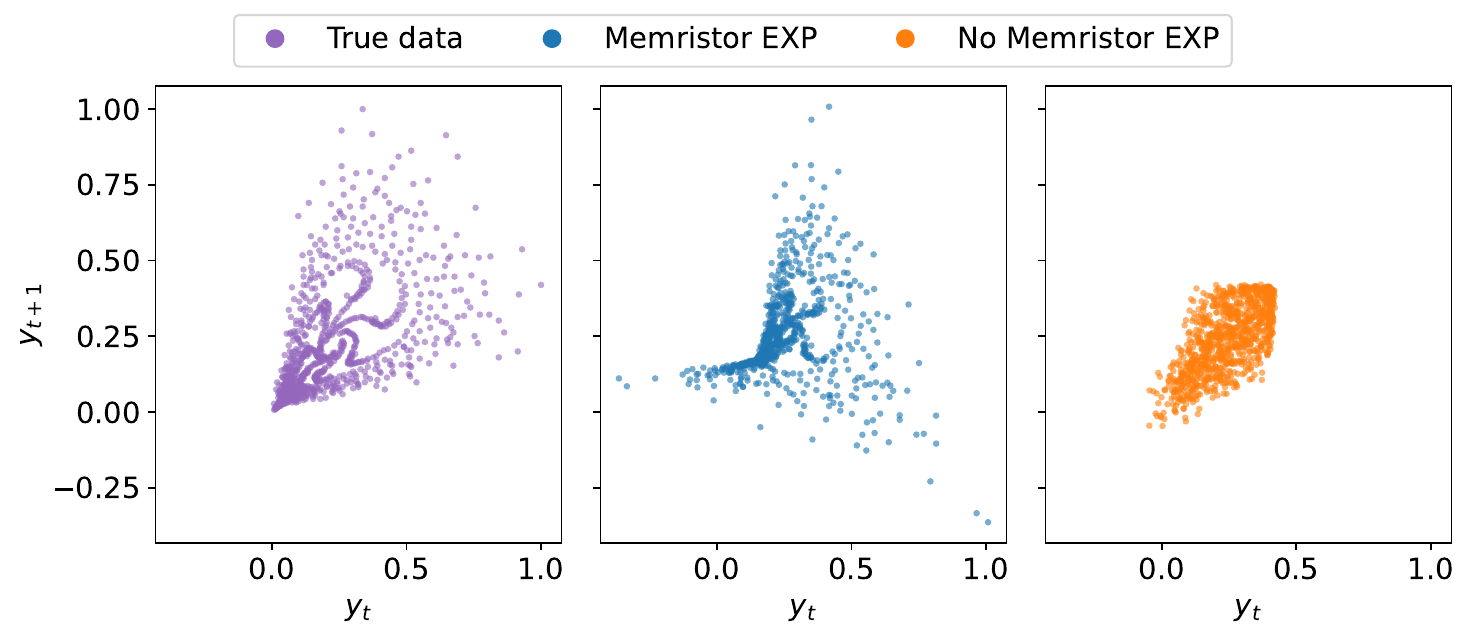}
        \vspace{1mm}
        (c)
    \end{minipage}

    \caption{\textbf{Lag plots of time series predictions.} Here we show the relationship between $x$ values separated by $\tau$ time steps (i.e. $x_t$ vs $x_{t+\tau}$) for the NARMA (a), Mackey-Glass (b) and SantaFe (c) tasks. The left panel shows the ground truth data, the middle panel displays the output from the experimental memristor, and the right panel corresponds to the model without the feedback loop. For NARMA and SantaFe tasks $\tau = 1$ and for Mackey-Glass task $\tau=10$.}
    \label{fig_lag}
\end{figure}

To test the effectivity of our model, we consider four tasks: the prediction of a nonlinear function and three different time series prediction. To highlight the difference between the case where we implement the feedback loop and the case where we do not do it, we report the lag plots of the three tasks in Fig.~\ref{fig_lag}.

\subsection{Nonlinear function reproduction/prediction}

For the first task, the functions we aim to reproduce are of the following kind:
\begin{equation}
    f(x) = x^n, \ n \in \mathbb{Z}
    \label{eq:f_2_predict}
\end{equation}
This choice is motivated by the interest in testing the achievable nonlinear behaviour of our model and the role of the feedback loop of the quantum memristor towards its enhancement.
 
We encode the inputs as  $x \rightarrow \sqrt{x}|0\rangle + \sqrt{1-x}|1\rangle$. To this end, we inject a photon from the lower input port of MZ1 of our photonic integrated platform (shown in Fig.~\ref{fig:SM-memristor}b) and use a unitary $U_{enc}$ to produce a coherent path superposition over the two modes A and B. In the Fock basis, this can be expressed as $x \rightarrow U_{enc}|0,1\rangle_{AB}= \sqrt{x}|1,0\rangle_{AB} + \sqrt{1-x}|0,1\rangle_{AB}$. In this notation, 1 indicates the presence of a photon and 0 the vacuum state. However, since we use the so-called dual-rail encoding, we take $|1,0\rangle_{AB}=|0\rangle$ and $|0,1\rangle_{AB}=|1\rangle$.

Let us note that this encoding is linear when performing a projective measurement in the eigenvector basis of the Pauli operator $\sigma_Z$ \cite{govia_nonlinear_2022}, as $Tr(\sigma_Z \rho_x) = \sum_{i=0,1}(-1)^ip_i=2x-1$, where $p_i$ is the probability of getting outcome $i$. On the contrary, a measurement in the $\sigma_X$ basis gives an expectation value that is nonlinear in $x$, $Tr(\sigma_X \rho_x) = \sqrt{x}\sqrt{1-x}$. This nonlinearity can be enhanced by rotating the state and choosing a proper measurement basis at the output. We indicate these two rotations $U_1$ and $U_2$ respectively, and perform them after the encoding (still through MZI1, which will implement $U_1 \cdot U_{enc}$) and before the measurement (through MZI3, which will implement $U_{meas} \cdot U_2$).

Then, to implement the quantum memristor, we add an ancillary spatial mode (C), and implement a unitary, $U_{mem}$, through MZI2, which acts on modes B and C (see Fig.~\ref{fig:SM-memristor}b). Subsequently, the feedback rule modifies $U_{mem}$, with the update rule of the form $a p_F +c$, where $p_F$ is the probability of detecting one photon coming out from output F.
This modification can be defined, when we model the quantum memristor as a beam splitter with tunable reflectivity $R$ (as in Fig.~\ref{fig:SM-memristor}a), as: 

\begin{equation}
    R_t=\frac{1}{m}\sum_{T=t-m}^t R_T=\frac{1}{m}\sum_{T=t-m}^t (a^*p_{FT}+b^*)
\end{equation}

where $m$ is the memory extent of the model, i.e. the number of previous outcomes which directly affect the value $R(t)$.
The relation between $R$ and the internal phase of MZI2 ($\theta$) is $R=cos^2(\theta/2)$.

We consider then the rotations $U_1$ and $U_2$, along with $a$ and $b$, as optimizable hyperparameters. More precisely, we minimize the following loss function: $p_{D} - x^{n}$
where $p_D$ is the probability of detecting one photon coming out from mode D of the interferometer. This is done numerically, by using a gradient descent algorithm  \cite{ruder2016overview}, in combination with \textit{Adam} optimizer \cite{diederik2014adam}, with a learning rate of 0.4. 

After finding the best hyperparameters, we run the experiment and feed the experimental statistics (i.e. a three-dimensional array containing the probabilities of detecting one photon on each output mode of the circuit) into the aforementioned artificial neural network, which is trained to perform the given task. We performed the task also without any feedback loop, keeping the $U_{mem}$ fixed, to compare the performance of the model in the two cases. 

This comparison, with and without the memristor feedback loop, makes it possible to distinguish the contributions of encoding and feedback loop to the overall nonlinearity. We note that to have a fair comparison, we used hyperparameters that were optimized separately for these two models. 

\subsection{Time-series prediction tasks}

We chose following three task to benchmark our quantum memristor: NARMA \cite{hochreiter_long_1997, atiya_new_2000}, Mackey-Glass \cite{shahi2022prediction,fujii_harnessing_2017} and SantaFe \cite{hubner1989dimensions, weigend1993results}.

\textbf{NARMA task.} The goal of this task is to predict the output of the following nonlinear dynamics \cite{atiya_new_2000, hochreiter_long_1997}:
\begin{equation}
    y_{t+1} = 0.4 y_t + 0.4 y_t y_{t-1} + 0.6x_t^3 + 0.1,
    \label{eq:narma}
\end{equation}
where $x_t$ and $y_t$ indicate, respectively, the input and the output at time $t$. The input of the time series $\bm{x} = \{x_i\}_{i=1}^N$ is generated by sampling each $x_i$ independently and uniformly in the interval $(0, \frac{1}{2})$ and, from Eq.~\eqref{eq:narma}, it is visible that the output $y_{t+1}$ depends on past two outputs $y_{t}$ and $y_{t-1}$ as well as input $x_t$.  

\textbf{Mackey-Glass task.} This task involves predicting the future evolution of the Mackey-Glass time series, which is a well-known example of a delayed nonlinear dynamical system. The Mackey-Glass equation is given by \cite{mackey1977oscillation}:  
\begin{equation}
    \frac{dx}{dt} = \beta \frac{x_{\tau}}{1 + x_{\tau}^n} - \gamma x,
    \label{eq:mackey-glass}
\end{equation}
where $x_{\tau} = x(t - \tau)$ introduces a time delay $\tau$, and the parameters $\beta$, $\gamma$, and $n$ control the system's behavior. For sufficiently large $\tau$, the system exhibits chaotic dynamics, making prediction a challenging task.  
To generate the dataset, we numerically integrate Eq.~\eqref{eq:mackey-glass} using Runge-Kutta method. The objective is to predict $x_{t+1}$ based on a history of past value(s), depending on the model used. The presence of delayed dependencies requires a reservoir computing model to exhibit both memory and nonlinearity to achieve accurate predictions.

\textbf{Santa Fe laser task.} The Santa Fe time-series prediction task is based on experimental data from a far-infrared laser operating in a chaotic regime. This dataset, originally introduced in the Santa Fe Time Series Competition \cite{weigend1993results}, presents a real-world example of chaotic dynamics, making it a widely used benchmark for reservoir computing models. The dataset consists of a univariate time series $\bm{x} = \{x_i\}_{i=1}^{N}$, where each $x_i$ represents the laser intensity at a discrete time step. The goal of the task is to predict the future value $x_{i+1}$ given a knowledge of past values, which depends on the model used. Due to the chaotic nature of the signal, a successful prediction requires the reservoir to capture both short-term correlations and the complex nonlinear structure of the underlying dynamics.

For all the tasks, we use a standard preprocessing procedure where the time series is scaled down, so the maximal value is one, and split into washout, training and testing sets. The washout corresponds to an initial amount of data which is required to properly initialize the model, as such, this is not considered in the prediction. This allows the final prediction to be independent on the initial values of the model.

\textbf{Feedback loop function.} Analogously to the previous task (see section A), an input sequence $\bm{x}$ is encoded into a quantum state, through a unitary $U_{enc}$ applied to a fixed state $|1\rangle$. In this case, the map is the following: $x_i \mapsto x_i \ket{0} + \sqrt{1-x_i^2}\ket{1}$ (for NARMA) and $x_i \mapsto \sqrt{1-x_i} \ket{0} + \sqrt{x_i}\ket{1}$ (for Mackey Glass and Santa Fe). Then, the feedback rule to update the memristor reflectivity was chosen to follow the recursive equation
\begin{equation}
    R_t = R_{t-1} + \frac{p_{t-1,2} - R_{t-1}}{m_d},
    \label{eq:ama}
\end{equation}
where $p_{ti}$ is the probability of detecting one photon in the $i$-th mode ($i=D,E,F$), at time $t$, while $m_d \geqslant 1$ is the \textit{memory decay} of the memristor, to which we also refer to as \textit{sliding window} (see section C for further details). Let us note that the feedback rule is linear and requires storing only two numbers in the classical memory, which are the previous reflectivity and the measurement outcomes. This update rule is called \textit{exponential moving average} and, with respect to the one used for the previous task (see section A), it only remembers one previous data point.

\begin{figure}[tb]
  \centering
  \includegraphics[width=0.9\textwidth]{  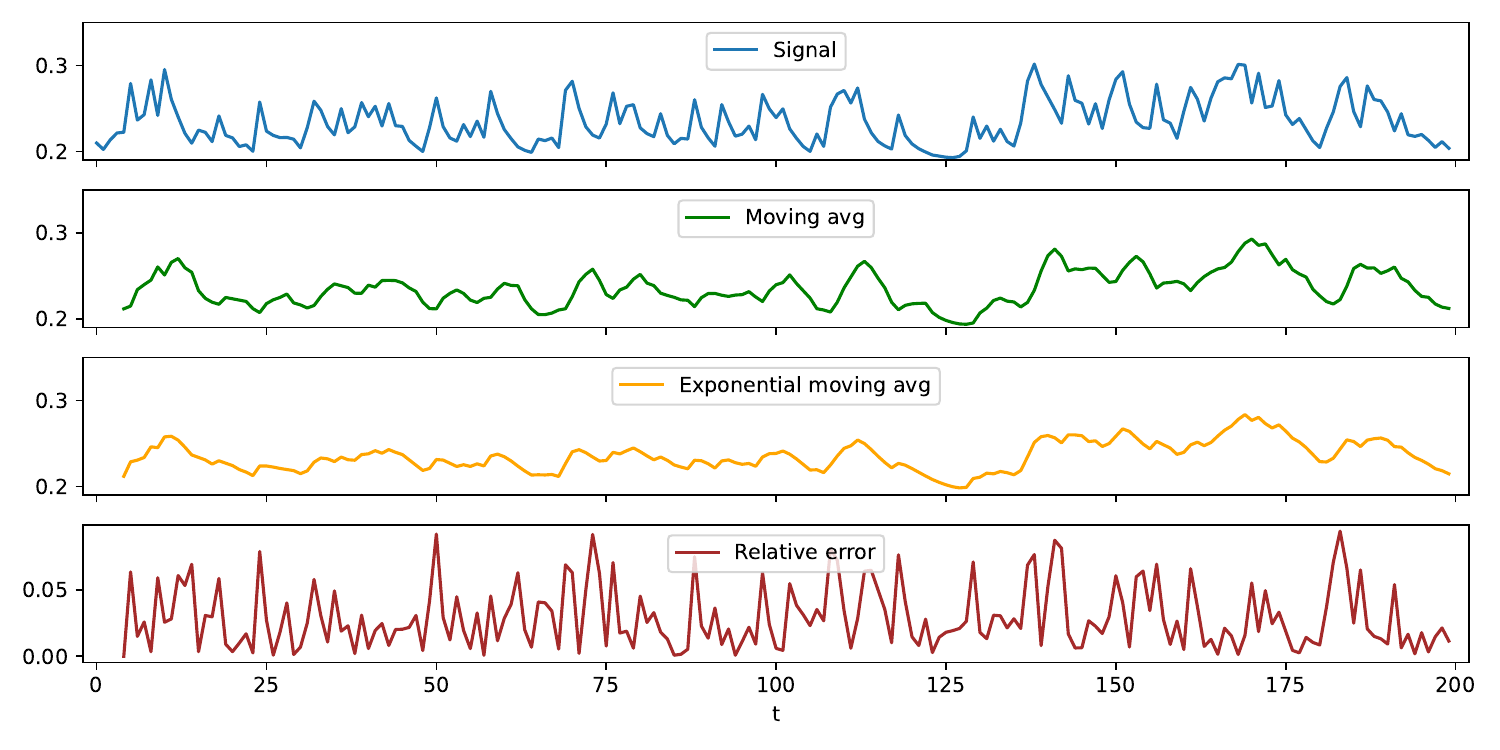}
  \caption{\textbf{Feedback rules comparison.} Moving average (green) and exponential moving average (orange) on the NARMA task (blue) with sliding window of 4. The red line indicates the relative error between two averages in each point, which never exceeds 10\%.}
  \label{fig:averages}
\end{figure}

The comparison of these two feedback rules, that is, the  moving average and exponential moving average, is shown in \cref{fig:averages}.

\begin{figure}[th]
    \centering
    \includegraphics[width=\linewidth]{  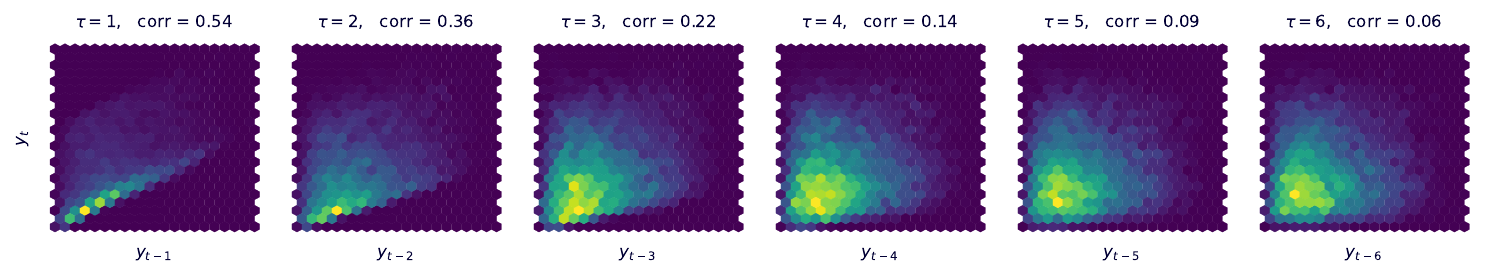} 
    \caption{\textbf{Consecutive output correlations in the NARMA task.} The plots show the correlations between $y_t$ and $y_{t-\tau}$ for different time values ($\tau$). The only significant correlations are up to $\tau = 3$. The numerical values for correlations at the top of each figure are Pearson correlation coefficient.}
    \label{fig:NARMA_correlations}
\end{figure}

Solving the recursion equation in \cref{eq:ama} yields:
\begin{equation}
    R_t = \left( 1 - \frac{1}{m}\right)^t \cdot R_0 + \frac{1}{m} \sum_{i=0}^{t-1} \left(1-\frac{1}{m}\right)^{t-1-i} 
    \cdot p_{i,2}, 
    \label{eq:mov_av_exp}
\end{equation}
where $R_0$ is the initial value of the reflectivity, which we set to $R_0 = \frac{1}{2}$ (balanced beam-splitter). Let us note that the name \textit{exponential moving average} comes from the fact that it amounts to a weighted moving average with exponentially decaying weights. Indeed, from Eq.~\eqref{eq:mov_av_exp}, we can note that the memory of the initial condition and of the probabilities that are too far in the past, decay exponentially. This is convenient for the NARMA task, where the correlations between $y_t$ and $y_{t-\tau}$ quickly disappear with increasing $\tau$, as shown in \cref{fig:NARMA_correlations}. In the extreme cases, i.e. for 
$m=1$ and $m \to \infty$, the memristor behaves as if it had no memory of previous inputs, i.e. $R_t = p_{t-1,F}$, or with fixed behaviour where $R_t \approx R_{t-1}$.

\begin{figure}[tb]
    \centering
    \includegraphics[width=.8\linewidth]{  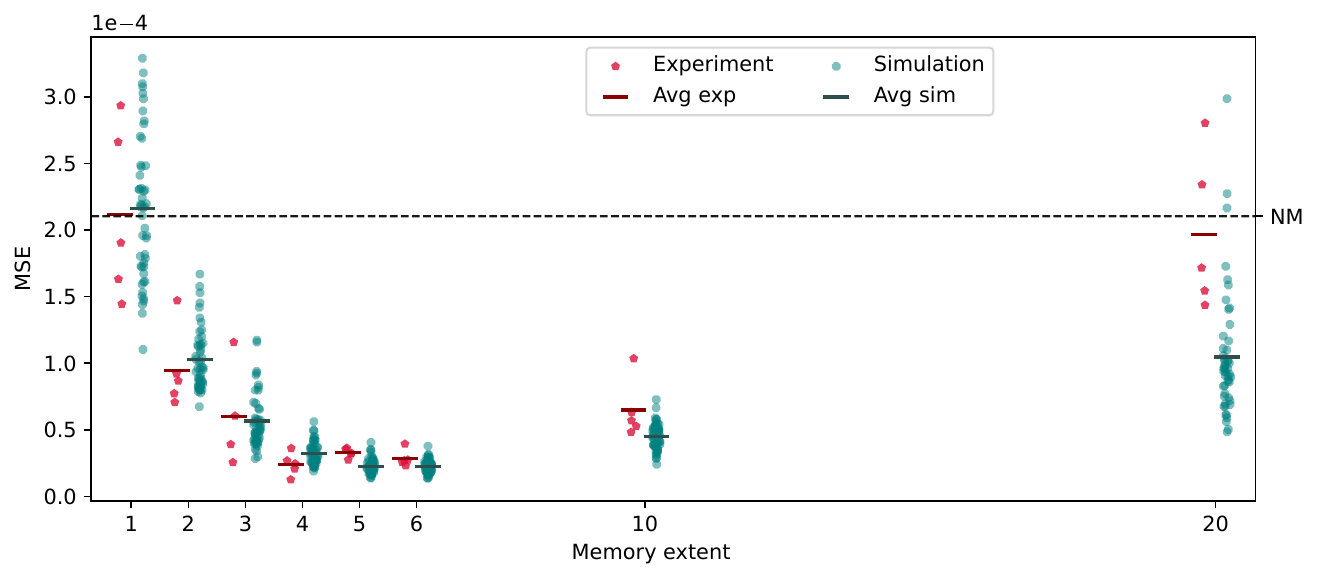}
    \caption{\textbf{Mean squared error (MSE) for different memory decays in NARMA task.} We report the MSEs obtained from 50 (numerical) and 5 (experimental) runs with green and red polygons, respectively. The dark red and dark green lines indicate the average values. The average MSE gets lower with the increasing size of the sliding window ($m$) and it reaches its minimum at $\sim 2.5 \times 10^{-5}$ for $m = 6$. The dashed line (NM) indicates the case with no feedback loop.}
    \label{fig:sliding_window_long}
\end{figure}

The optimal trade-off value for $m$ strongly depends on the task and can be treated as an hyperparameter. In our case, we chose $m=4$, following the behaviour depicted in~\cref{fig:sliding_window_long}.

In \cref{fig:sliding_window_long}, we show the performance of the model for different sliding window parameters. Let us note that large values worsen the prediction, both in the numerical calculations and the experimental data. Indeed, for $N=20$ the prediction from the experiment tends to be, on average, as good as the model without memory.

\subsection{Comparison with classical models} 

In order to benchmark the performance of our quantum memristor-based reservoir computing scheme, we chose some models to compare it with. In detail, we picked two classes: 1) models without memory; 2) models with the memory of the previous output.

\textbf{Models without memory.} Those try to predict the next value of the time series ($y_{t+1}$) using \textbf{only} the current value ($y_t)$. The equation of this model is 
\[
    y_{t+1} \approx \text{Poly}_d(y_t),
\]
where $\text{Poly}_d(\cdot)$ is the polynomial of the degree at most $d$. We consider two models of this kind: linear and cubic. We can write them explicitly:
\[
\begin{aligned}
    y_{t+1} &\approx \text{Poly}_1(y_t) = \alpha y_t + \beta  
    &\quad &\text{for linear model}, \\
    y_{t+1} &\approx \text{Poly}_3(y_t) = \alpha y_t^3 + \beta y_t^2 + \gamma y_t + \delta 
    &\quad &\text{for cubic model}.
\end{aligned}
\]
The optimal coefficients, denoted by Greek letters, are determined using linear regression. Let us note that these models have, respectively, 2, 4, 3 and 9 free parameters to be optimized.

\textbf{Models with memory.} Here we discuss the model with the simplest memory possible, which is a polynomial access to the current the previous output. The equation of the model is 
\[
    y_{t+1} \approx \text{Poly}_d(y_t, y_{t-1}),
\]
where $\text{Poly}_d(\cdot)$ is the polynomial of the degree at most $d$. We consider two models of this kind: linear and cubic. We can write them explicitly:
\[
\begin{aligned}
    y_{t+1} &\approx \text{Poly}_1(y_t, y_{t-1}) = \alpha y_t + \beta y_{t-1} + \gamma 
    &\quad &\text{for linear model}, \\
    y_{t+1} &\approx \text{Poly}_3(y_t,  y_{t-1}) = \alpha y_t^3 + \beta y_t^2 y_{t-1} + \gamma y_t y_{t-1}^2 + \delta y_t^2 + \epsilon y_t y_{t-1} + \zeta y_{t-1}^2 + \eta y_t + \theta y_{t-1} + \iota
    &\quad &\text{for cubic model}.
\end{aligned}
\]
The optimal model parameters, denoted by Greek letters, are determined using linear regression.

\begin{figure}[p]
    \centering
    
    \begin{minipage}{\columnwidth}
        \centering
        \includegraphics[width=0.65\columnwidth]{  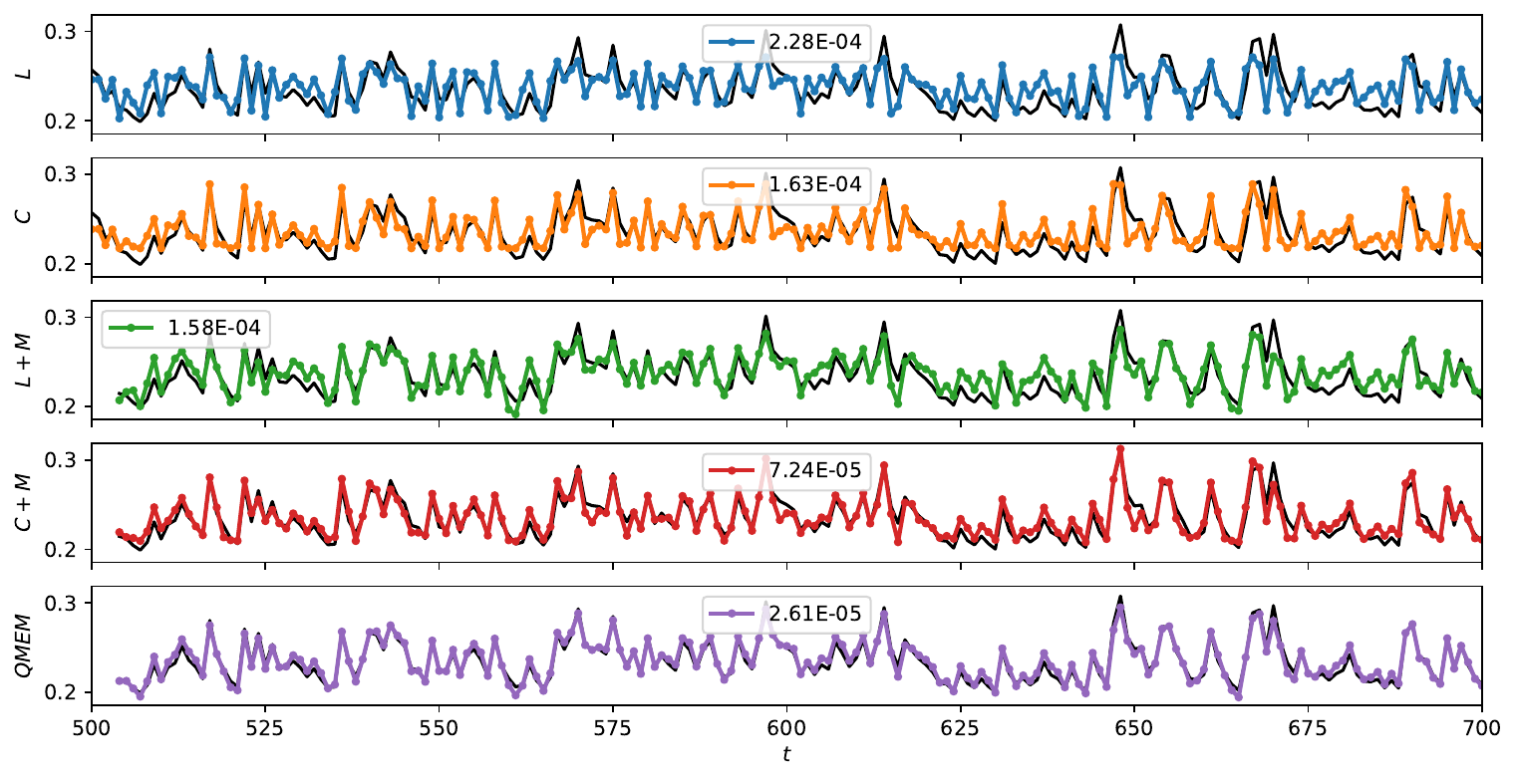}
        \vspace{1mm}
        (a)
    \end{minipage}
    
    \vspace{4mm}
    
    \begin{minipage}{\columnwidth}
        \centering
        \includegraphics[width=0.65\columnwidth]{  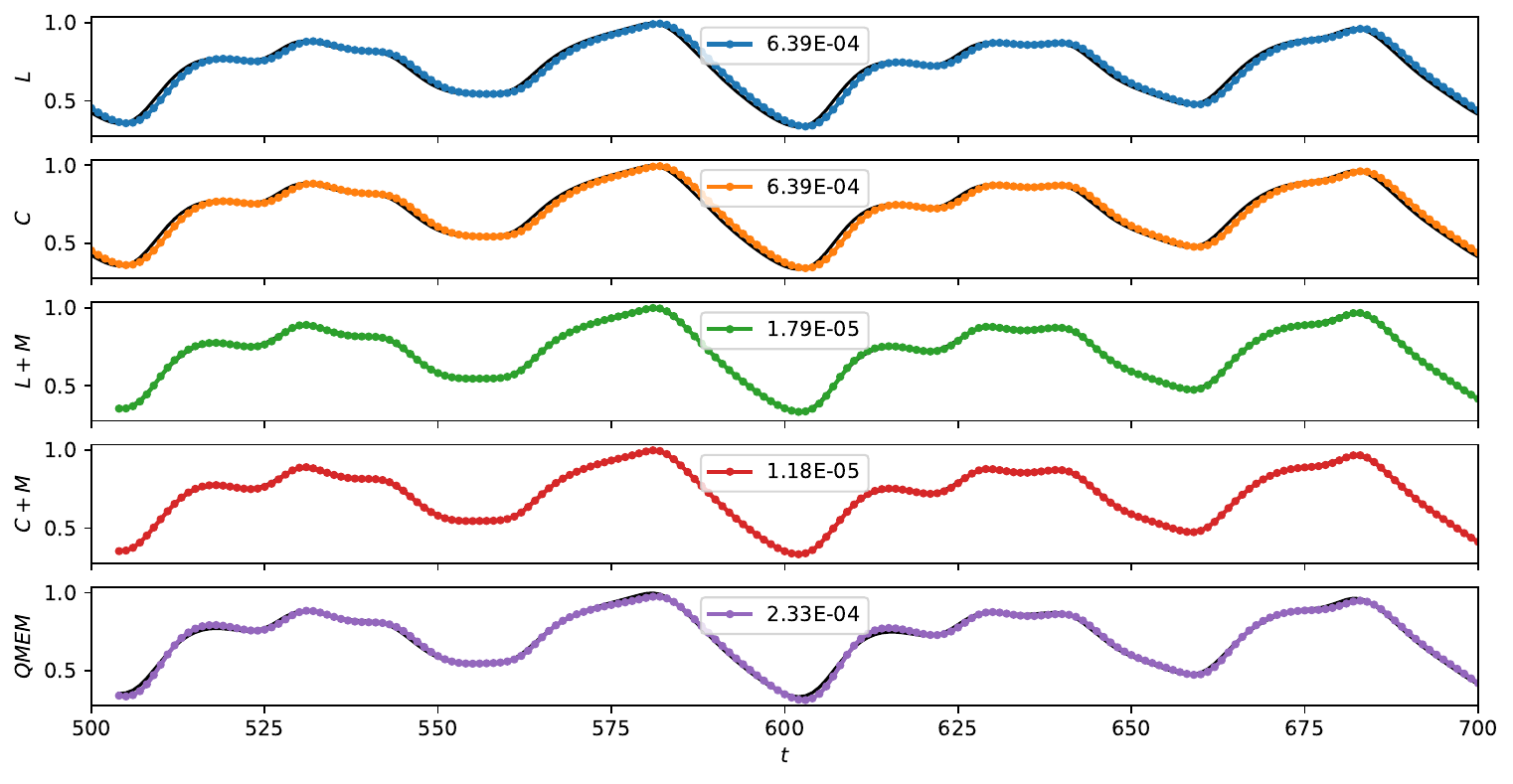}
        \vspace{1mm}
        (b)
    \end{minipage}
    
    \vspace{4mm}
    
    \begin{minipage}{\columnwidth}
        \centering
        \includegraphics[width=0.65\columnwidth]{  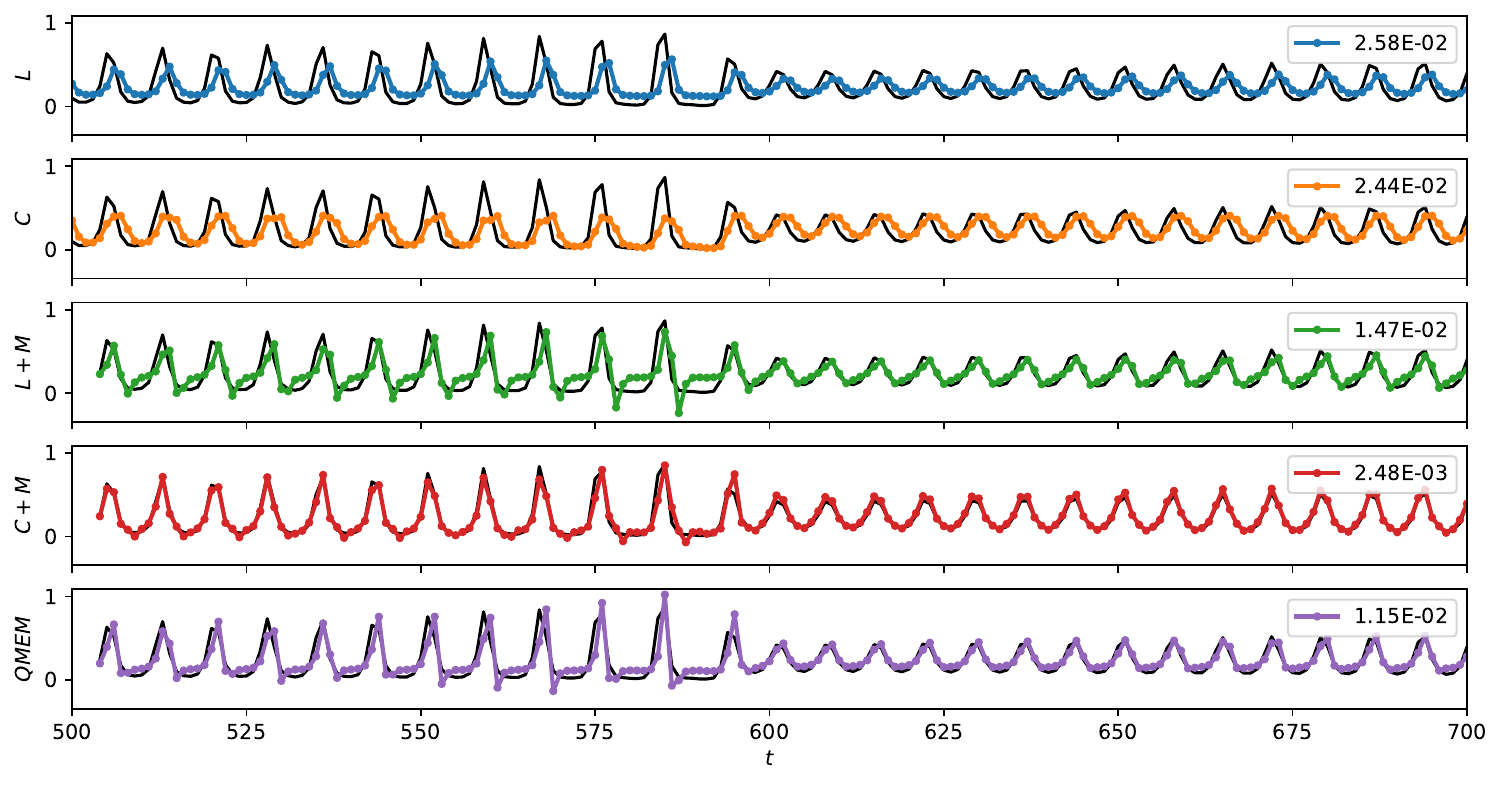}
        \vspace{1mm}
        (c)
    \end{minipage}

    \caption{\textbf{Performance of different learning models on time-series prediction tasks.} The plot shows the prediction of the chosen two classes of classical models on the NARMA (a), Mackey-Glass (b) and SantaFe (c) task compared to the true values of the time series (solid black line). For every plot, the quality of the prediction is quantified by the mean-squared error (MSE), which is reported in the inset on the upper right. In detail, $L$ and $C$ indicate a model with no memory which can manipulate the input, respectively, linearly and cubically. $L+M$ and $C+M$ indicate a model which can manipulate the input, respectively, linearly and cubically and have access to the previous input of the model. Finally, $QMEM$ indicates the model with a quantum memristor with a feedback loop. All models are trained on the same dataset of length 1000. The first 20 data points are used to washout the memory, the following 480 points for training and the remaining 500 for testing. For the sake of clarity, we display only 200 points from the test dataset.}
    \label{fig:model-comparison}
\end{figure}

In both memory and no memory cases and in all three tasks, we noticed that the prediction does not improve when $d>3$, thus  we chose $d=3$ to present the results, for all the tasks. Details are presented in the Fig. \ref{fig:model-comparison}.

\begin{figure}[h]
    \centering
    \includegraphics[width=0.5\linewidth]{  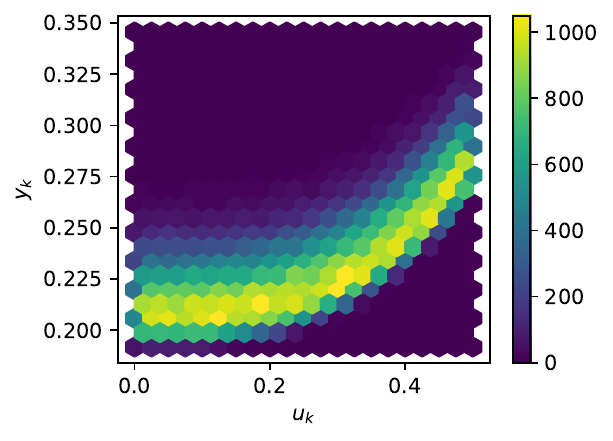}
    \caption{\textbf{Input/output relation for NARMA task}. The plot shows that the nonlinearity of the NARMA task is not very high, for inputs between 0 and 0.5. This explains why also the considered linear models have reasonably good performances on this task.}
    \label{fig:u_y_correlation}
\end{figure}

We notice that for NARMA task the model with a quantum memristor performs the better than to classical models presented earlier. Let us note that also linear models with limited memory have a good performance, due to the low nonlinearity of the NARMA task, which is highlighted in \cref{fig:u_y_correlation}. For the Mackey-Glass task, we see that the quantum model not a significantly better performance than the classical models. The explanation lies in the smoothness of the function to predict, as well as in the fact that probe points are dense, making the function locally flat. Lastly, in the SantaFe task quantum model is better than the models without the memory and is comparable to the model with a simple memory. However it is outperformed by the cubic model with memory. The reason lies in the number of free parameters that can be optimized in the classical case. The overall conclusion that we can draw from this comparison is that the quantum model seems to be superior for time series which are not smooth and cannot be locally approximated by a polynomial of a small degree. However it always performed better than the models without access to previous outputs, highlighting the fact that the feedback loop successfully implements a memory of previous states.

\begin{table}[h]
    \renewcommand{\arraystretch}{1.5} 
    \centering
    \begin{tabular}{c||c|c||c|c}
        \textbf{MSE} &
        \shortstack{Experiment \\ with \\ memristor} & 
        \shortstack{Simulation \\ with \\ memristor} & 
        \shortstack{Experiment \\ without \\ memristor} & 
        \shortstack{Simulation \\ without \\ memristor} \\ \hline
        \hline
        $f(x)=x^4$ & $5.7 \times 10^{-4} $ & $7.1 \times 10^{-4}$ & $1.67 \times 10^{-2}$ & $2.3 \times 10^{-2}$ \\ 
        NARMA & $2.8 \times 10^{-5} $ & $2.6 \times 10^{-5}$ & $2.3 \times 10^{-4}$ & $2.0 \times 10^{-4}$ \\ 
        Mackey--Glass & $2.4\times 10^{-4}$ & $2.2 \times 10^{-4}$ & $9.4 \times 10^{-4}$ & $6.4 \times 10^{-4}$ \\ 
        SantaFe & $23\times 10^{-3}$ & $9.2\times 10^{-3}$ & $2.8\times 10^{-2}$ & $2.5\times 10^{-2}$\\ 
    \end{tabular}
    \caption{\textbf{Experimental and numerical achieved MSE scores, with and without the memristor.} The experimental imperfection have little effect on the NARMA and Mackey--Glass tasks, but it significantly impacts the Santa Fe task. The average is performed on 3 independent runs.}
    \label{tab:exp_sim_compar}
\end{table}

\begin{figure}[tb]
    \centering
    \includegraphics[width=0.7\linewidth]{  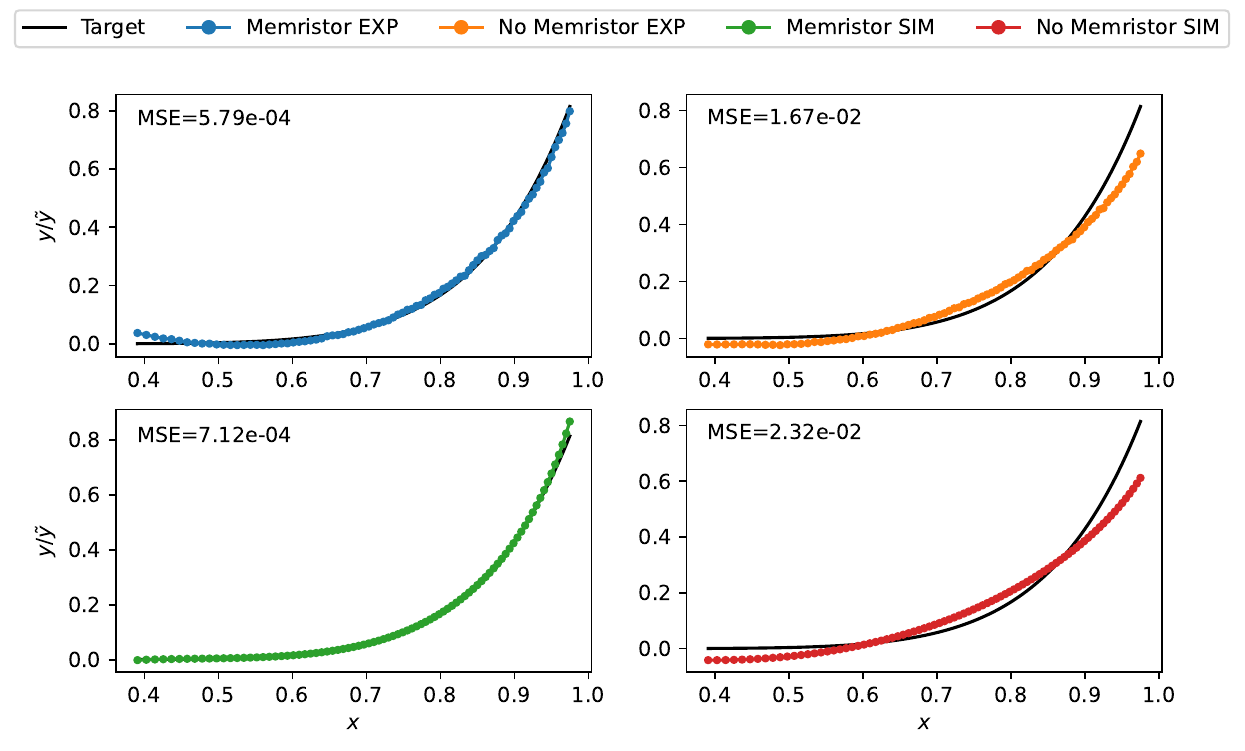}
 \caption{\textbf{Experimental results vs numerical simulations for nonlinear function prediction.} These plots compare the experimental results with simulation results for the nonlinear function prediction of $x^4$. The blue and green lines represent the experimental and simulation results, respectively, for the case with the memristor. In contrast, the orange and red lines correspond to the same task without a memristor, i.e., with no feedback loop, again showing the experimental and simulation results, respectively. The quality of the prediction is quantified by the mean-squared error reported in the plots. This comparison demonstrates that the simulation and experimental results are in good agreement.}
    \label{fig:exp_vs_th_P}
\end{figure}

\begin{figure}[tb]
    \centering

    \begin{minipage}{\columnwidth}
        \centering
        \includegraphics[width=0.7\columnwidth]{  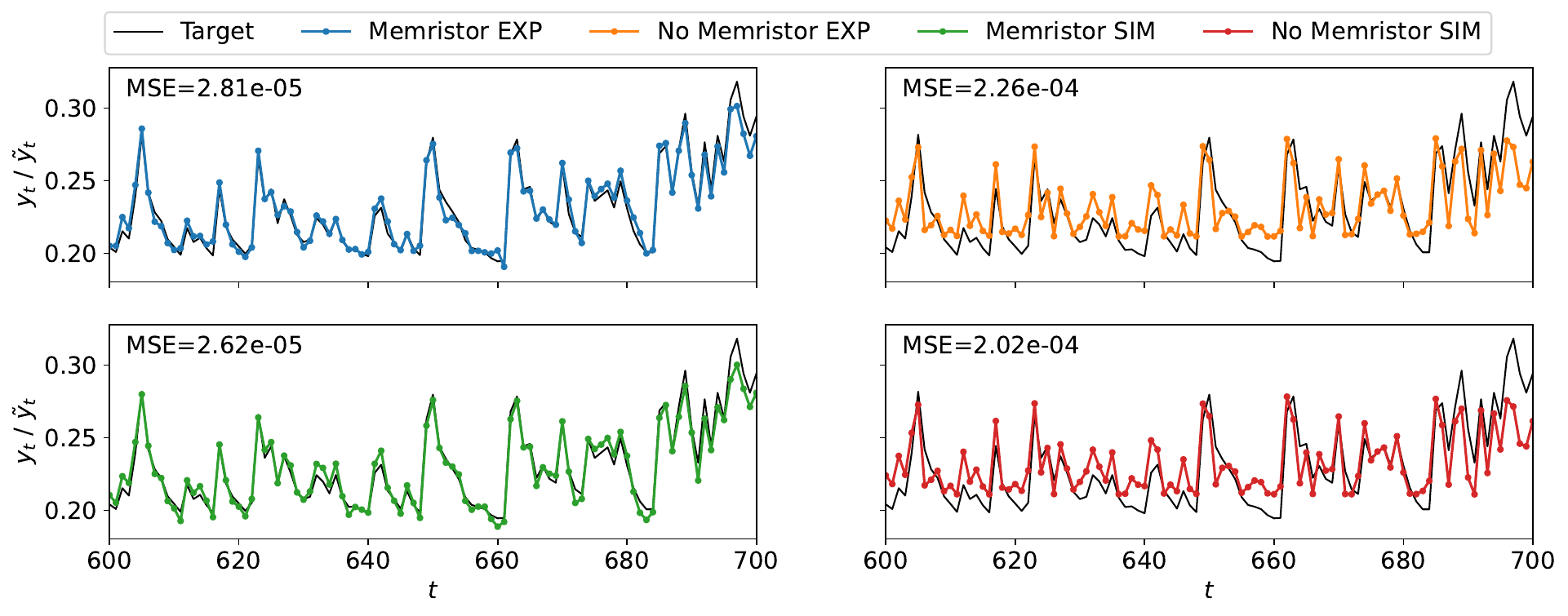}
        \vspace{1mm}
        (a)
    \end{minipage}

    \vspace{4mm}

    \begin{minipage}{\columnwidth}
        \centering
        \includegraphics[width=0.7\columnwidth]{  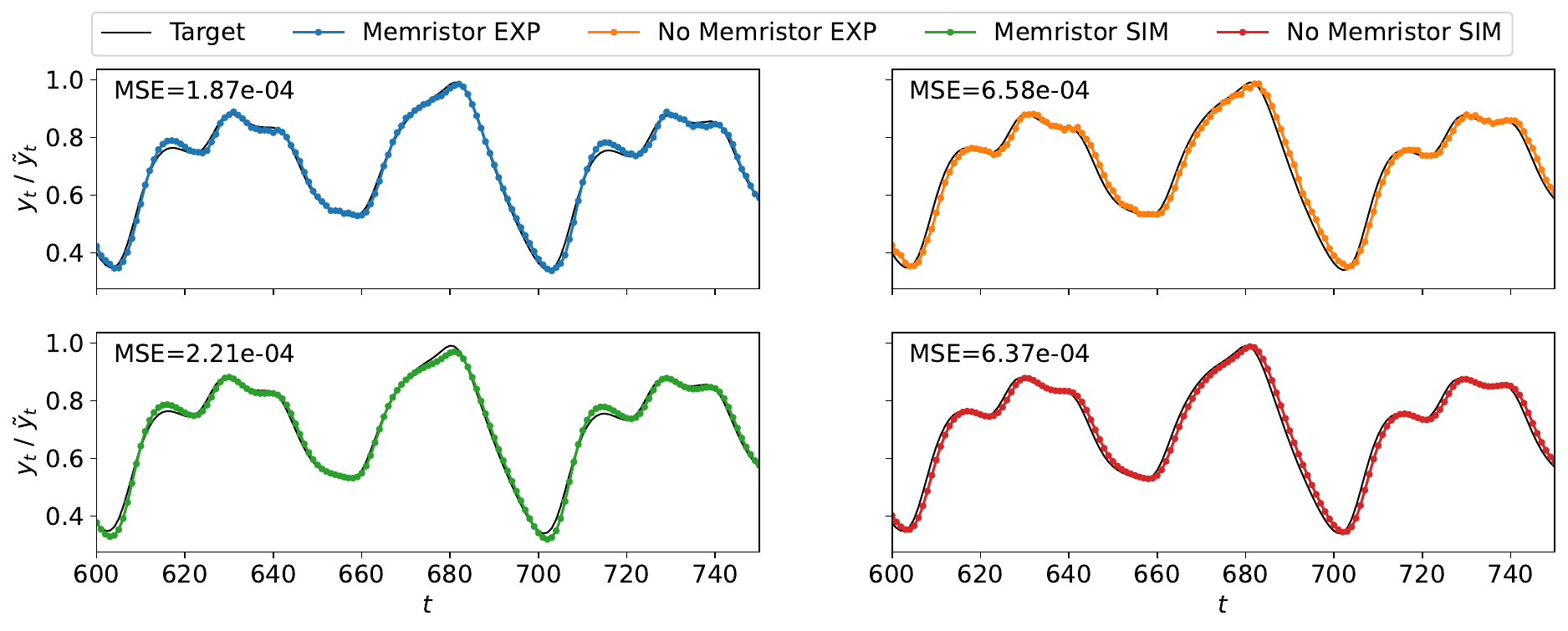}
        \vspace{1mm}
        (b)
    \end{minipage}

    \vspace{4mm}

    \begin{minipage}{\columnwidth}
        \centering
        \includegraphics[width=0.7\columnwidth]{  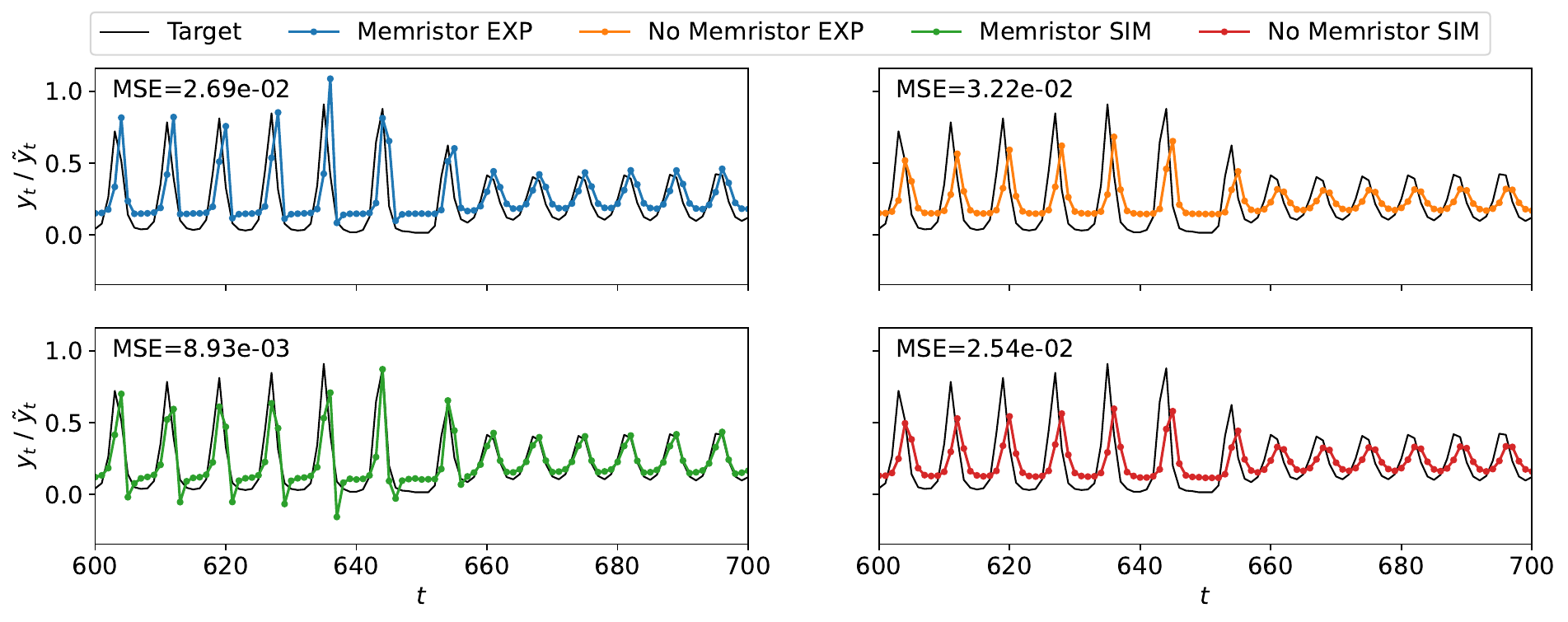}
        \vspace{1mm}
        (c)
    \end{minipage}

    \caption{\textbf{Comparison of experimental results with classical simulations for (a) the NARMA task, (b) Mackey-Glass time series prediction, and (c) Santa Fe time series prediction.} The blue and green lines represent cases with a memristor in the experimental and numerical scenarios, respectively, while the orange and red lines correspond to cases without a feedback loop in the experimental and numerical scenarios. The time series consists of 1000 data points, divided into \textit{washout} (20), training (480), and testing (500) phases. To improve readability, only 100 points from the test dataset are shown. The Mean Squared Error (MSE) scores for each plot are displayed in the top-left corner.}
    \label{fig:exp_sim_compar}
\end{figure}

\begin{figure}[ht]
    \centering
    \includegraphics[width=0.4\linewidth]{  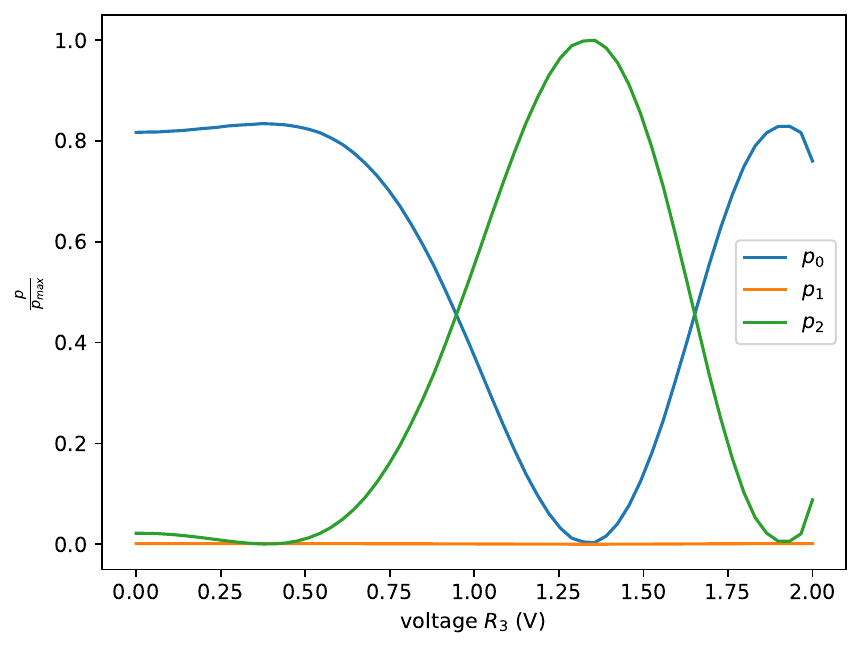}
    \caption{\textbf{Calibration of phase shifter corresponding to resistor $R_3$.} The optical power is measured at the output modes D, E and F, i.e. $p_0$, $p_1$, and $p_2$ respectively while applying a range of voltages to the phase shifting resistor $R_3$. The depicted measured optical output power values for each voltage setting are fitted to a polynomial fit that allows for precise mapping from voltage to phase.}
    \label{fig:characterisation-R3}
\end{figure}

\subsection{Comparison between experiment and numerical simulations}

To benchmark the control that we have over our experimental apparatus and understand the noise robustness of our model, we provide here a comparison between the experimental results presented in this work, along with numerical simulations using the same datasets. 
Regarding the monomial predictions, we report the comparison in Fig.~\ref{fig:exp_vs_th_P}.

Then, for the other tasks, Table \ref{tab:exp_sim_compar} shows the MSE obtained by averaging over 5 repetitions for all tasks and combinations, both with and without the memristor. We observe that the experimental and numerical predictions for the NARMA and Mackey–Glass datasets show a good agreement. Instead, the Santa Fe dataset is more affected by noise, suggesting that the impact of noise is task-dependent. Additionally, Fig.~ \ref{fig:exp_sim_compar} presents visual comparisons for all combinations of experiment vs. numerical simulation and with vs. without the memristor. It is important to note that the MSE score in the figures corresponds to the individual dataset score, not the average value shown in Table \ref{tab:exp_sim_compar}.

\section{Calibration of on-chip phase shifters}
Before conducting any experiments, it is essential to characterize the resistive elements within our circuit to ensure accurate phase control. In general, the characterization process involves applying a range of voltages to the phase-shifting resistors and measuring the resulting phase shifts. By injecting light into a specific mode and varying the applied voltage, we can determine how each resistor modulates the phase.

We begin with the characterization of the third resistor, $R_3$ (see Fig.~\ref{fig:SM-memristor}), as it can be characterized independently by injecting light into the last mode (C) of the integrated circuit. At this step, we apply no voltage on the other phase shifters. A Mach-Zehnder interferometer transforms the input state $|\psi_{\text{in}}\rangle$ through the unitary transformation $\textbf{U}_{\text{MZI}}$,
\[
|\psi_{\text{out}}\rangle = \textbf{U}_{\text{MZI}} |\psi_{\text{in}}\rangle
\]

with

\[
\textbf{U}_{\text{MZI}}
=
\begin{pmatrix}
\sin \frac{\theta}{2} & \cos \frac{\theta}{2} \\
\cos \frac{\theta}{2} & -\sin \frac{\theta}{2}
\end{pmatrix}
\]
where $\theta$ is the internal phase.
This allows us to compute the expected output state $|\psi_{\text{out}}\rangle$ given a known input. At the output modes (D, E and F), we measure the optical powers ($p_0$, $p_1$ and $p_2$). When characterizing $R_3$, since we do not know the passive phases implemented by MZI1 and MZI3, we need to conclude the following:

\[
 p_0+p_1 = |\langle \psi_1 | \psi_{\text{out}} \rangle|^2 = \left| \sin \frac{\theta}{2} \right|^2 P_{max}
\]

and 

\[
 p_2 = |\langle \psi_1 | \psi_{\text{out}} \rangle|^2 = \left| \sin \frac{\theta}{2} \right|^2 P_{max}
\]

where $\langle \psi_1 |=\langle 10|$ and $P_{max}$ is the total power injected in the circuit. Indeed, the optical power coming out of MZI2 on the upper mode can be randomly split by MZI3.
By fitting these measured values using a polynomial fit, we establish a precise mapping from voltage to phase $\theta$. In Fig.~\ref{fig:characterisation-R3}, we show the experimental values for the normalized intensity at each output port of the interferometer.

Once $R_3$ is characterized, we can proceed to the first resistor, $R_1$. To do so, we set the phase of $R_3$ to $0$, to reroute the fraction of light that exits the lower mode of MZI1, to the output mode F. With this configuration, we inject light from the second mode (B) and again vary the voltage applied to $R_1$, measuring the corresponding output powers to extract the phase response. In this case, the sum of the optical powers fractions coming out from modes $D$ and $E$ will correspond to $p_0+p_1=|cos(\frac{\theta}{2})|^2$ and the one coming out from F to $p_2=|sin(\frac{\theta}{2})|^2$.

Finally, we characterize the fifth resistor, $R_5$. At this stage, we configure $R_1$ and $R_3$ such that their respective transformations act as identity matrices. We then inject light and vary the voltage on $R_5$, following the same procedure of measuring the output power and fitting the data to obtain the voltage-to-phase mapping.
In this case, we will have $p_0=|sin(\frac{\theta}{2})|^2$, $p_1=|sin(\frac{\theta}{2})|^2$ and $p_2=0$.

The characterization of the external resistors of the MZIs, $R_2$ and $R_4$, cannot be performed independently. Here we set $R_1$ and $R_5$ to act as balanced beam splitters and $R_3$ to act as identity. Effectively, this configuration creates an MZI where the total internal phase is given by the sum of $R_2$ and $R_4$.
